\documentclass{nature}

\usepackage{color}
\usepackage{bm}
\usepackage{graphicx}
\usepackage{lineno}
\usepackage{CJKutf8}
\usepackage{soul}
\usepackage{pdfpages}

\DeclareMathAlphabet{\mathbfit}{OML}{cmm}{b}{it}
\makeatletter
\let\saved@includegraphics\includegraphics
\AtBeginDocument{\let\includegraphics\saved@includegraphics}
\renewenvironment*{figure}{\@float{figure}}{\end@float}
\makeatother

\title{\begin{center}
Triplon current generation in solids
\end{center}
}

\author{ Yao Chen$^{1}$, Masahiro Sato$^{2,\dagger}$, Yifei Tang$^1$,  Yuki Shiomi$^{3}$, Koichi Oyanagi$^{1,4}$,
 Takatsugu Masuda$^{5}$,  Yusuke Nambu$^{1,6,7}$, Masaki Fujita$^{1}$,  \& Eiji Saitoh$^{1,8,9,10,11,*}$}

\begin{document}

\maketitle

\begin{affiliations}
 \item Institute for Materials Research, Tohoku University, Sendai 980-8577, Japan
 \item Department of Physics, Ibaraki University, Mito, Ibaraki 310-8512, Japan
\item Department of Basic Science, The University of Tokyo, Tokyo 153-8902, Japan
 \item Faculty of Science and Engineering, Iwate University, Morioka 020-8551, Japan
 \item Institute of Solid State Physics, The University of Tokyo, Kashiwa, Chiba 277-8581 Japan
 \item FOREST, Japan Science and Technology Agency, Kawaguchi, Saitama 332-0012, Japan
 \item Organization for Advanced Studies, Tohoku University, Sendai 980-8577, Japan
 \item Department of Applied Physics, The University of Tokyo, Tokyo 113-8656, Japan
 \item Institute for AI and Beyond, The University of Tokyo, Tokyo 113-8656, Japan
 \item Advanced Institute for Materials Research, Tohoku University, Sendai 980-8577, Japan
 \item Advanced Science Research Center, Japan Atomic Energy Agency, Tokai 319-1195, Japan 
 \item[] * e-mail address: eizi@ap.t.u-tokyo.ac.jp
 \item[] $\dagger$ e-mail address: masahiro.sato.phys@vc.ibaraki.ac.jp
\end{affiliations}

\begin{abstract}
\par

\section*{Abstract}

A triplon refers to a fictitious particle that carries angular momentum $\mbox{\boldmath $S$ = 1}$ 
corresponding to the 
elementary excitation in a broad class of quantum dimerized spin systems.
Such systems without magnetic order have long been studied as a testing ground for quantum properties of spins. 
Although triplons have been found to play a central role in thermal and magnetic properties in dimerized magnets with singlet correlation, a spin angular momentum flow carried by triplons, a triplon current, has not been detected yet. 
Here we report spin Seebeck effects induced by a triplon current: triplon spin Seebeck effect, using a spin-Peierls system 
 CuGeO$_3$.
The result shows that the heating-driven triplon transport induces spin current 
whose sign is positive, opposite to the spin-wave cases in magnets. 
The triplon spin Seebeck effect persists far below the spin-Peierls transition temperature, 
being consistent with a theoretical calculation for triplon spin Seebeck effects.

\end{abstract}

\par 

\section*{Introduction}

Spin Seebeck effects\cite{Uchida:2008cc} (SSEs) refer to the generation of a spin current, a flow of spin angular momentum of electrons, from a temperature gradient applied to a bilayer system comprising a magnet and heavy metal such as Pt.
Spin current in the magnet propagates along the temperature gradient and reaches the heavy metal\cite{Takahashi:2010ju}.
When the randomly polarized electrons in the  heavy metal scatter at the interface with magnet, the electron spin experience an exchange interaction form the spin current carrier of the magnet. Such interaction aligns spins in the heavy metal along the spin direction of spin current carrier in the magnet and induces spin accumulation at the heavy metal side of interface. The diffusion of such spin accumulation causes a pure spin current in the heavy metal, in which spin current can be detected as a voltage signal via the inverse spin-Hall effect (ISHE)\cite{Saitoh:2006kk,Valenzuela:2006cs}.
The SSE has been observed in various insulating systems, in which parasitic thermal effects originating from itinerant electrons can be eliminated. Some types of spin carriers
 including magnons in magnets\cite{Ito:2019bq,2015NatPh..11.1022C,Seki:2015esa,Wu:2016hh}, paramagnons\cite{Wu:2015gf}, antiferromagnetic magnons\cite{Seki:2015esa,Wu:2016hh}, and spinons in a spin liquid\cite{Hirobe:2016eg,Hirobe:2018ds} have been demonstrated for different mechanisms of the SSE.

\par 

Among quantum spin systems without magnetic order, dimerized magnets spin systems occupy an important position, in which two neighboring spins are frozen as $S=0$ singlets in the ground state.
The elementary spin excitations are $S=1$ triplet states, called a triplon. 
A triplon has been expected to carry spin angular momentum.

\par 
In this work, we report the observation of spin current in CuGeO$_3$, carried by triplons in terms of longitudinal SSE measurements. A typical spin-dimer system is a spin-Peierls material
 CuGeO$_3$, which contains 
 one-dimensonal spin-1/2 chains with antiferromagnetic exchange interaction for nearest-neighbor spins.
In the spin chain, with lowering the temperature, neighboring spins dimerizes to form a spin-gapped phase;
the transition is called a spin Peierls (SP) transition\cite{Pytte:1974jt}.
Below the SP transition temperature $T_{\mathrm{SP}}$, the chain
distorts so that the distances between neighboring spins change alternately.  
The bond-alternating exchange interaction causes neighboring spins to dimerize to reduce the total energy, creating a gap in the spin excitation energy spectrum. CuGeO$_3$ is the firstly discovered inorganic system that exhibits a SP transition\cite{Hase:1993bo}. 
High-quality single crystals of CuGeO$_3$ are easier to obtain than organic SP materials and CuGeO$_3$ has typically been used to study spin excitations in SP systems\cite{2002JPCM...14R.195U}.

\par 
\section*{Results}

\subsection{Sample structure and the concept of the study.} 
Figure 1a shows the crystal structure of CuGeO$_3$. 
One dimensional Cu$^{2+}$ spin ($S=1/2$) chains are aligned along the $c$-axis, as illustrated as dotted lines. 
The nearest-neighbor exchange interaction along the $c$-axis is $J_c \sim 120 $ K, much greater than the interchain exchange coupling\cite{Nishi:1994da}, $J_b\sim 0.1 J_c$ and $J_a\sim -0.01 J_c$ along the $b$ and $a$-axis, respectively. 
CuGeO$_3$ is thus considered as a quasi-one-dimensional spin system.
  As the temperature $T$ decreases below the SP transition temperature  $T_{\mathrm{SP}} \sim  14.5$ K, the lattice of CuGeO$_3$ is spontaneously distorted to form a bond alternating configuration\cite{Lorenzo:1994bt}.
  The spin configuration in the ground state is shown schematically in Fig. 1b, where all spins are frozen as dimerized $S=0$ singlets.
The elementary spin excitation from the ground state is an $S=1$ triplon~\cite{Takayoshi:2010fi}, also shown in Fig. 1b. 
  The excitation gap of a triplon is estimated to be $\sim 23$ K from ESR\cite{Brill:1994ki} and neutron scattering\cite{Regnault:1996bv,Fujita:1995ha,Nishi:1994da} experiments.

\par 

Based on the previous neutron scattering results\cite{Regnault:1996bv} and theoretical calculations\cite{Bonner:1982bs},
the dispersion relation of triplons along the $c$-axis is sketched in Fig. 1c.
In the absence of a magnetic field, $\mu_0 H = 0$, the triplon states are threefold degenerated:
$\left|\uparrow \uparrow \right\rangle$, $\Bigl(\left|\uparrow\downarrow\right\rangle+\left|\downarrow\uparrow\right\rangle\Bigr)/\sqrt{2}$, and $\left|\downarrow \downarrow \right\rangle$ corresponding to states with different spin quantum numbers, $S_z = 1, 0$, and $-1$, respectively.
 When a magnetic field is applied, the degenerated triplet bands split into three due to the Zeeman effect, where the triplet state with the
  parallel spin direction to $H$, $\left|\uparrow \uparrow \right\rangle$, has the lowest energy\cite{Brill:1994ki,Regnault:1996bv,Fujita:1995ha}.
   The $\left|\uparrow \uparrow \right\rangle$ state thus exhibits the highest probability for thermal excitation, while $\left|\downarrow \downarrow \right\rangle$ the lowest.
Driven by a thermal gradient, different occupancy among the three states can result in a flow of a net spin angular momentum in an external magnetic field: triplon SSE. 
One of the key features of the triplon SSE is that it should exhibit the opposite sign of ISHE signal to the magnon SSE in ferri- and ferro- magnets.
This is because triplon spin currents are carried by excitation with the parallel spin direction to the external field; while magnons carry antiparallel spins to the external field, as shown in Fig. 1d.
Due to the opposite polarization direction of spin current carrier in these two systems, the electron spin in Pt feels reverse direction of exchange interaction, resulting in spin accumulation with reverse polarization direction and different sign in ISHE. 
Our theoretical calculation based on the tunnel spin current (see Supplementary Note G3 for details) suggests that the opposite sign of triplon SSE to magnon SSE is a universal property for gapped triplet-spin systems.

\par 
\subsection{Material characterization and measurement setup.}
First, we measured temperature dependence of magnetization, $M(T)$, for the CuGeO$_3$ sample.
Figure 1e shows the magnetic susceptibility $\chi(T) = M(T)/H$ measured at $\mu_0H = 1$ T $\parallel$ $b$-axis. 
When $T$ is decreased from 20 K, $\chi$ exhibits a sudden drop at the SP transition temperature $T_{\mathrm{SP}} \sim 14.5 $ K and rapidly decreases towards zero. 
 This indicates that Cu spins along the spin chain ($\parallel c$-axis) form dimers and a spin gap develops below $T_{\mathrm{SP}}$.
By measuring $\chi(T)$ at different values of $H$ (see Supplementary Fig. 1 for details), we obtain the $H$ - $T$ spin-phase diagram of the CuGeO$_3$ sample in Fig. 1f.
In the phase diagram, for large $H$ in the SP phase, CuGeO$_3$ 
undergoes another phase transition into a magnetic phase at a transition field, $H_{\mathrm{m}}$, where magnetization is recovered by forming a spin soliton lattice in the chain\cite{Ronnow:2000dt}.
The transition can be confirmed as a sudden jump of magnetization as a function of fields\cite{Hase:1994ez} (also see Supplementary Fig. 1).
The obtained phase diagram is in good agreement with a previous study\cite{Uchinokura:1995cu}.

\par 

The experimental setup for the SSE measurement and the sample used in the present study are shown in Figs. 2a and 2b.
A trilayer structure comprising a Pt wire for ISHE, an insulating SiO$_2$ layer, and an Au wire as a heater was fabricated on the top of the $(001)$ plane of a CuGeO$_3$ single crystal.
  The spin Seebeck voltage signal $V_{\mathrm{SSE}}$ in the Pt wire is measured by a lock-in method, where an a.c. current is applied to the Au layer under an in-plane magnetic field, $H$, applied perpendicular to the Pt wire.
  We normalize $V_{\mathrm{SSE}}$ by the heater power and the detector resistance as $\tilde{V}_{\mathrm{SSE}}=V_{\mathrm{SSE}}/(P_{\mathrm{Au}}\cdot R_{\mathrm{Pt}})$.

\par 

\subsection{Observation of the triplon SSE.} 
The magnetic field dependence of $\tilde{V}_{\mathrm{SSE}}$ at some selected temperatures is shown in Figs. 2c and 2d. 
At $15$ K, just above $ T_{\mathrm{SP}} \sim 14.5 $ K, no $H$-dependent signal is recognized.
Most notably, a clear voltage signal appears when $T$ is decreased down to $2$ K. 
The signal is an odd function of $H$, reflecting the symmetry of the ISHE\cite{maekawa2012spin}. 
The sign of the $\tilde{V}_{\mathrm{SSE}}$ signal is opposite to that of the magnon-mediated SSE\cite{Kikkawa:2015bn,2010ApPhL..97q2505U}, which is confirmed in a similar setup for a ferrimagnetic insulator Y$_3$Fe$_5$O$_{12}$/Pt sample in Figs. 2e and 2f.
The result indicates that spin current carriers in these two systems have different spin polarization directions with respect to $H$, consistent with the triplon-current senario.
 Note that the sign of the SSE signals in spin nematic LiCuVO$_4$/Pt   (ref.~\cite{Hirobe:2019cf}) is also opposite to the present CuGeO$_3$/Pt case.
 We also note that, unlike the magnon-mediated SSE in ferro/ferrimagnets, the triplon-mediated SSE dose not scale with the $M(H)$ curve. For the magnon-mediated SSE\cite{2010ApPhL..97q2505U}, $M(H)$ represents the increase in the spin-current polarization. 
  In contrast, the magnetization increase in the SP phase means larger density of broken dimers and the scattering probability between triplons and broken dimers becomes lager.
  The unusual $H$-dependence of the triplon SSE originates from the different properties between ferro/ferrimagnets and SP phases, and it is one of the characteristics of triplon SSE. 
 (see Supplementary Fig. 2 for details).

\par 

 As shown in Figs. 3a and 3b, the temperature dependence of $\tilde{V}_{\mathrm{SSE}}$ shows that the signal magnitude increases monotonically as $T$ decreases from $T_{\mathrm{SP}}$ to $3$ K. When $T < 3$ K, the SSE voltage remains almost unchanged down to $2$ K.
 At higher temperatures, the dimerized ground state is weakened because the thermally excited phonon disturbs the spontaneous displacement of the Cu$^{2+}$ ion. 
 In other words, the singlet ground state and triplon excitation is no longer a good picture for the low-energy excitation, and the triplon lifetime shrinks toward higher temperatures and completely disappears at $T_{\mathrm{SP}}$.
 This effect has been observed as an increase in the peak width of the triplon excitation in neutron scattering experiments\cite{Lussier:1996gja}. 
We note that, by phenomenologically incorporating the triplon-phonon scattering term in a calculation, the $T$ dependence of $\tilde{V}_{\mathrm{SSE}}$ at moderate temperatures can be explained qualitatively (see Supplementary Note G4 for details).

\subsection{Influence of crystal orientation and Zn-doping on the triplon SSE.}  
Let us turn to the $H$ dependence of $\tilde{V}_{\mathrm{SSE}}$ in CuGeO$_3$/Pt.
In the entire temperature range, as $H$ increases from zero, $\tilde{V}_{\mathrm{SSE}}$ increases first and takes its maximum at $\mu_0H \sim  2 $ T. 
As $H$ further increases, $\tilde{V}_{\mathrm{SSE}}$ is then suppressed.
As shown theoretically in the following, triplon scattering by inevitable impurities and defects in the nominally pure CuGeO$_3$ sample can explain the observed $H$ dependence of $\tilde{V}_{\mathrm{SSE}}$.

\par 
  
To show that the observed voltage is related to the triplon in the SP phase, we performed another control experiment with $\nabla T$ applied perpendicular to the spin chain configuration.
Figure 3c shows $\tilde{V}_{\mathrm{SSE}}(H)$ for $\nabla T \perp c$ in a reference sample prepared with the same procedure as $\nabla T \parallel c$. 
The observed SSE signal in $\nabla T \perp c$ 
is only $1/7$ in magnitude as compared to the $\nabla T \parallel c$ configuration, consistent 
with the highly anisotropic transport of triplons. This result corroborates that the origin of the observed signal is the 1-D triplon transport. Moreover, 
additional reference experiments are performed for two nonmagnetic Zn-doped CuGeO$_3$ samples, 
in which the triplon spin current should be suppressed by Zn-induced scattering.
Figures 3d and 3e show the $\tilde{V}_{\mathrm{SSE}}(H)$ for Cu$_{0.99}$Zn$_{0.01}$GeO$_3$/Pt and Cu$_{0.97}$Zn$_{0.03}$GeO$_3$/Pt in the same temperature range as the CuGeO$_3$/Pt measurements. 
Clearly, $\tilde{V}_{\mathrm{SSE}}$ is suppressed in the $1\%$ Zn-doped sample and disappeared in the $3\%$ Zn-doped sample.

\par 

Figures 3f and 3g show $\chi$ for Cu$_{0.99}$Zn$_{0.01}$GeO$_3$ and Cu$_{0.97}$Zn$_{0.03}$GeO$_3$, respectively. 
For Cu$_{0.99}$Zn$_{0.01}$GeO$_3$, the SP transition is observed at 13.2 K. The decrease of $T_{\mathrm{SP}}$ in doped CuGeO$_3$ agrees with the result from the previous reports\cite{Sasago:1996is}. 
On the other hand, the SP transition is not visible in $\chi(T)$ in Cu$_{0.97}$Zn$_{0.03}$GeO$_3$.
Instead, a clear cusp is observed in $\chi_{c}(T)$ (with $H\parallel c$-axis) at $T_{\mathrm{N}}\sim 4.3$ K, indicating phase transition to an antiferromagnetic phase. 
The $T_{\mathrm{N}}$ is consistent with the previously reported value     for Zn-$3\%$ doped CuGeO$_3$ (ref.~\cite{Sasago:1996is}).
The doping effect implies that non-magnetic Zn breaks the Cu-spin chains and introduces unpaired Cu free spins, called solitons, into the spin chain.
Solitons disturb the dimerization, and local antiferromagnetic correlation may lead to long-range antiferromagnetic order\cite{Fukuyama:1996hb} at low temperatures in Cu$_{0.97}$Zn$_{0.03}$GeO$_3$. 
As shown in Fig. 3h, $\tilde{V}_{\mathrm{SSE}}$ is significantly suppressed in the $\nabla T \perp$ spin-chain configuration and the Zn-doped CuGeO$_3$ samples in all $T$ range. 
In the doped samples, the spin-excitation gap is still present in the SP phase, but their transport may be blocked by Zn. 
The result suggests that the triplon spin current is strongly influenced by Zn-induced scattering. 
Curie-Weiss fitting of $\chi(T)$ at low $T$ ($< 3$ K)\cite{Hase:1993bo,Grenier:1998ga}(see Supplementary Note F for details) reveals that the density of free $S=1/2$ spins is estimated to be $0.02\%$ in the nominally pure CuGeO$_3$ sample, so the effect of triplon scattering by impurities cannot be ignored.
Therefore, we attempted to formulate the triplon spin current in the manner of Boltzmann equation, which is capable of incorporating scattering effects.

\par 
\subsection{Theoretical model for the triplon SSE.}  
Figure 4 compares the $H$ dependence of the observed SSE voltage signal with  theoretical results. 
As we already discussed, $|\tilde{V}_{\mathrm{SSE}}|$ is expected to monotonically grow with increasing $H$ 
because the triplon density of $S_z=1$, i.e., the dominant carrier of spin current, increases with lowering the spin gap 
up to the critical field $H_{\mathrm{m}}$. 
However, the observed broad peak (see Fig. 4a) and the Zn-doped results shown in Fig. 3 strongly suggest 
the importance of triplon scattering processes. 
To theoretically analyze triplon currents in $\rm CuGeO_3$, 
we here apply the Boltzmann equation\cite{abrikosov2017fundamentals}.
The spin current $J_s$ computed by the Boltzmann equation (See Supplementary Note G2) 
is given by
\begin{equation}
	J_s  = \sum _{S_z} \int \frac{k}{2\pi} \hbar S_z (v_{S_z}(k))^2\tau_k k_{\mathrm{B}}\Delta T\frac{\epsilon_{S_z}(k)}{(k_{\mathrm{B}} T)^2} \frac{\exp(\epsilon_{S_z}(k)/k_{\mathrm{B}}T)}{(\exp\left[\epsilon_{S_z}(k)/k_{\mathrm{B}}T\right]- 1)^2},
\label{eq:spincurrent}
\end{equation}
where $\Delta T$ is the temperature difference along the spin chain direction. 
$v_{S_z}(k)$, $\tau_k$ and $\epsilon_{S_z}(k)$ are the group velocity, relaxation time, and energy dispersion 
for the triplons\cite{Takayoshi:2010fi}, respectively. We assume that the triplon can be viewed as a bosonic particle and 
the final part of Equation (\ref{eq:spincurrent}) stems from the Bose distribution function.
The $S_z = 0$ branch of the triplon does not contribute to the spin current. 
At $H=0$, the distribution functions of $S_z = \pm 1$ branches are identical, which cancel out the total spin current. 
By varying $T$ and $H$, the thermal excitation and the Zeeman effect alter the triplon distributions. 
Assuming a constant $\tau_{k}$, the population difference between $S_z = \pm 1$ increases monotonically 
at a fixed $T (<T_{\rm SP})$ due to the Zeeman effect, resulting in an almost $H$-linear increase in $J_s$. 
The $H$-linear behavior can also be reproduced by using 
a tunnel spin-current formula,\cite{Adachi:2011hh,Hirobe:2016eg,Hirobe:2019cf,Jauho:1994cg} 
which is another microscopic theory for thermal spin current (see Supplementary Note G3). 

\par 
As shown in Fig. 4a, $|\tilde{V}_{\mathrm{SSE}}(H)|$ takes a maximum at around 3 T. 
To explain the non-monotonic behavior of $\tilde{V}_{\mathrm{SSE}}$, we considered two mechanisms 
of triplon scattering processes. 
One is the scattering by nonmagnetic impurities, where the scattering probability 
$1/\tau_{k,\mathrm{non-mag}}$ is constant with respect to $H$.
Another source of scattering is free spins of broken dimers and unpaired Cu$^{2+}$ (S = $1/2$).
The density of scatterers and scattering potential are assumed to depend on $H$ as a Brillouin function. 
The total scattering probability is then the sum of the two scattering mechanisms, 
$1/\tau_{k,\mathrm{total}} = 1/\tau_{k,\mathrm{non-mag}} + 1/\tau_{k,\mathrm{mag}}$. 
Assuming that the scattering effect by magnetic impurities is dominant, 
we can reproduce the experimentally obtained $H$ dependence of  $\tilde{V}_{\mathrm{SSE}}$ shown in Fig. 4b. 
In the low $H$ regime, where the spin of magnetic impurities is not aligned and the scattering effect is weak, 
the Zeeman effect dominates the generation and transport of the triplon current. 
With increasing $H$, the triplon mode of $S_z = +1$ ($S_z = -1$) shifts downward (upward) due to the Zeeman effect, and the imbalance between $S_z = \pm 1$ triplons becomes larger, resulting in an increase in the total triplon current.
As $H$ is further increased, the impurities are magnetized and magnetic scattering dominates the triplon transport, resulting in a broad peak as shown in Fig. 4b.
The result agrees with the experimentally obtained data shown in Fig. 4a in a semi-quantitative level.

\par 

\section*{Discussion}
Finally, we here examine other possible origins for the observed thermoelectric voltage. 
One candidate could be the paramagnetic SSE due to free Cu spins. 
Several papers have addressed the spin current transport in paramagnets, including Gd$_3$Ga$_5$O$_{12}$ (refs.~\cite{Wu:2015gf,Oyanagi:2019cv}), DyScO$_3$ (ref.~\cite{Wu:2015gf}) and La$_2$NiMnO$_6$ (ref.~\cite{Shiomi:2014gd}). 
Paramagnon is believed to be responsible for spin transport in paramagnets as a result of short-range magnetic correlation or long-range dipole interactions.
 The spin diffusion length of Gd$_3$Ga$_5$O$_{12}$ is estimated to be about 1.8 $\mathrm{\mu m}$ at $5$ K\cite{Oyanagi:2019cv}. 
 However, in our CuGeO$_3$ sample with the free spin density of $0.02\%$, the average distance between free spins is estimated to be around $0.29$ nm$/0.02\% \sim\ 1.5\ \mathrm{\mu m}$ (the distance between neighboring Cu$^{2+}$ ions along the $c$-axis is $0.29$ nm\cite{Kamimura:1994fe}).
  This means that the paramagnetic spins are too dilute for spin correlation and spin current transport. 
Furthermore, the suppression of $V_{\mathrm{SSE}}$ in the 
Zn-doped samples, which exhibit larger paramagnetic moments than CuGeO$_3$, also rules out the possibility of paramagnetic SSE.

\par 

The normal and anomalous Nernst effects may also contribute to an electric voltage under a temperature gradient, but these effects contradict with the suppression 
of the signal in the high temperature spin-liquid phase; the vanishing signals in the $\nabla T \perp c$ configuration and the Zn-doped samples (see Figs. 3a-3e).
Furthermore, the normal Nernst effect should be linear with respect to the applied magnetic field and can not explain the observed magnetic field dependence of $V_{\mathrm{SSE}}$ shown in Figs. 3a and 3b.

\par 

Note that the voltage observed at 15 K in the CuGeO$_3$/Pt shows almost no $H$-dependence, as shown in Fig. 2c. 
Above $T_{\mathrm{SP}}$, the elementary excitation of spin chains are gapless spinons\cite{Bonner:1982bs}. 
A spinon spin current, a spin current carried by spinons, was found in Sr$_2$CuO$_3$ (ref.~\cite{Hirobe:2016eg}).
The spinon spin current may contribute to a small $H$-linear voltage signal above $T_{\mathrm{SP}}$ (see Supplementary Note C for details). In the spin-Peierls phase, the gapless spinon is replaced by the gapped triplon\cite{Bonner:1982bs}.

\par 

In summary, we observed the triplon spin-Seebeck effect in CuGeO$_3$/Pt.
Due to the triplon excitation in the system, 
the sign of the observed SSE is opposite to that of the conventional magnon SSE. 
The triplon SSE signal persists down to $2$ K in the spin-Peierls phase and
  the magnetic field dependence of the triplon SSE is consistent with microscopic calculation.
Our result shows that the spin-Seebeck effect also acts as a probe for spin excitations in gapped spin systems, and can also be applied to other materials with exotic spin excitation, such as a spin ladder system SrCu$_2$O$_3$ (ref. \cite{Azuma:1994hf}) and a spin dimer system SrCu$_2$(BO$_3$)$_2$ (ref. \cite{Kageyama:1999ke}).

\par 

\begin{methods}

\subsection{Sample fabrication.} 
CuGeO$_3$ single crystals were grown by a traveling solvent floating zone (TSFZ) method. Samples were cut into size of $\sim$ 7 mm $\times$ 3 mm for further measurements.
40nm-thick ferrimagnetic insulator Y$_3$Fe$_5$O$_{12}$ single crystals were grown on Gd$_3$Ga$_5$O$_{12} (111)$ substrates by magnetron sputtering.
Prior to the deposition, Gd$_3$Ga$_5$O$_{12}$ substrates were annealed in the air at 825 $^{\circ}$C for 30 min in a face-to-face configuration.
To crystalize the as-grown amorphous Y$_3$Fe$_5$O$_{12}$, the samples were post-annealed in air at 825 $^{\circ}$C for 200 s.
For SSE measurements, on-chip devices with Pt detector and SiO$_2$/Au heater were fabricated on top of samples (both pure/doped CuGeO$_3$ and Y$_3$Fe$_5$O$_{12}$) by an electron-beam lithography, magnetron sputtering and lift-off technique (see Supplementary Note A).

\subsection{ SSE measurement.}
We adapted the on-chip heating method for SSE, as illustrated in Fig.1b. 
 The resistance between the Au and Pt layers is much greater than $\mathrm{M\Omega}$ at room temperature. 
A longitudinal temperature gradient was generated by applying an a.c. current with the frequency of $f = 13$ Hz to the Au heater. 
Since the SSE is driven by Joule heating, the frequency of the SSE signal on the Pt wire is $2f$.
 The SSE signal was detected via a lock-in method.

\subsection{Magnetization measurement.} The magnetization of CuGeO$_3$ was measured in a Quantum Design Physical Properties Measurement System (PPMS) with a vibrating sample magnetometer (VSM) option.

\end{methods}

\begin{addendum}
 \item
 This is a post-peer-review, pre-copyedit version of an article published in Nature
Communications. The final authenticated version is available online at: 
 https://doi.org/10.1038/s41467-021-25494-7.
 We thank D. Hirobe, T. Kikkawa, G. E. W. Bauer and Y. Hirayama for valuable discussions. 
 This research was supported by JST ERATO Spin Quantum Rectification Project (No. JPMJER1402), JPSP KAKENHI (
 the Grant-in-Aid for Scientific Research (S) No. JP19H05600 and JP17H06137;
 the Grant-in-Aid for Scientific Research (A) No. JP16H02125;
 the Grant-in-Aid for Scientific Research (B) No. JP20H01830 and No. JP19H02424; 
the Grant-in-Aid for Scientific Research (C) No. JP17K05513;
the Grant-in-Aid for Challenging Research (Exploratory) No. JP19K22124 and No. JP20K20896;
the Grant-in-Aid for Scientific Research on Innovative Areas (Research in a proposed research area) No. JP20H04631, No. JP20H05153, No. JP19H05825 and No. JP19H04683;
the Grant-in-Aid for Research Activity start-up No. JP20K22476), JST CREST (No. JPMJCR20C1 and No. JPMJCR20T2) and MEXT (Innovative Area ``Nano Spin Conversion Science'' (No. 26103005)). Y.C., Y.T. and K.O. are supported by GP-Spin at Tohoku University. Y.C. is also supported by the Japan Society for Promotion of Science through a research fellowship for young scientists (No. JP18J21304). 
 \item[Author contributions] 
 Y.C. designed the experiments in discuss with Y.S. and E.S. Y.T., Y.N., T.M. and M.F. grew single crystals for this study. Y.C. and K.O. nano-fabricated devices. Y.C. collected and analyzed the experimental data. M.S. preformed the theoretical calculations. E.S. supervised this work. Y.C., M.S., Y.S. and E.S. wrote the manuscript. All authors discussed the results and commented on the manuscript.
  \item[Competing Interests] The authors declare that they have no competing interests.
\item[Data availability] The data that support the findings of this study are available from the corresponding
authors upon reasonable request.
\item[Code availability]
The codes used in theoretical simulations and calculations are available from the corresponding authors upon reasonable request.

\item[Correspondence] Correspondence and requests for materials
should be addressed to Y.C (e-mail:\newline y.chen.imr@gmail.com) and M.S (e-mail:masahiro.sato.phys@vc.ibaraki.ac.jp).
\end{addendum}

\newpage
\thispagestyle{empty}
\begin{figure}
\begin{center}
\includegraphics[width=17cm, bb= 0 0 256 141]{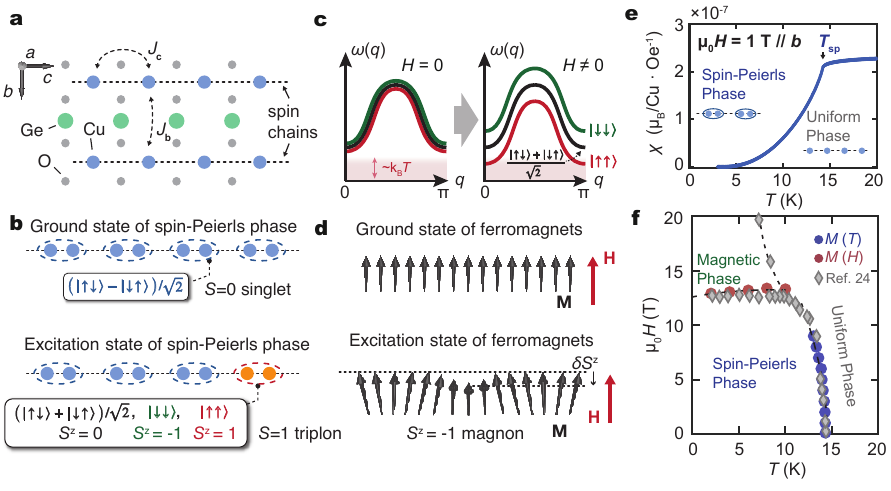}
\label{fig1}
\end{center}
\end{figure}

\newpage
\thispagestyle{empty}
\begin{figure}
\begin{center}
\caption{\textbf{Spin chain and spin-Peierls (SP) phase of CuGeO$_3$.} 
\textbf{a,} Crystalline structure of CuGeO$_3$. The Cu spin chains are shown as dotted lines.  
Spin-$1/2$ spin chains along the $c$-axis in CuGeO$_3$ are formed by staking CuO$_2$ chains. 
$J_c$ and $J_b$ denote the exchange interactions  along the $c$ and $b$-axis, respectively.
\textbf{b,} When temperature ($T$) falls down to the SP transition temperature $T_{\mathrm{SP}}$, CuGeO$_3$ undergoes a SP transition.
After the transition, the Cu spins dimerize and the ground state becomes a spin singlet state 
($\Bigl(\left|\uparrow\downarrow\right\rangle-\left|\downarrow\uparrow\right\rangle\Bigr)/\sqrt{2}$ with $S = 0$).
Elementary excitation in the SP phase is a triplon with $S=1$.
\textbf{c,} Schematic band structure of triplon in CuGeO$_3$\cite{Regnault:1996bv,Bonner:1982bs}. Three triplet states ($\left|\downarrow\downarrow\right\rangle$, $\Bigl(\left|\uparrow\downarrow\right\rangle+\left|\downarrow\uparrow\right\rangle\Bigr)/\sqrt{2}$ and $\left|\uparrow\uparrow\right\rangle$) are threefold degenerated at zero field.
By applying a magnetic field ($H$), the Zeeman effect lift the degeneracy.
\textbf{d,} The ground state and excitation state of a ferromagnet. Elementary excitation is a magnon with $S=1$. The magnons reduce the magnetization ($\mathbf{M}$) along the magnetic field ($\mathbf{H}$) as $\delta S^z$.
\textbf{e,} $T$ dependence of magnetic susceptibility ($\chi$) of CuGeO$_3$ under an applied magnetic field of $1$ T along the $b$-axis. The SP transition temperature $T_{\mathrm{SP}}$ is illustrated as the arrow in the figure.
\textbf{f,} Magnetic field -  temperature ($ H $ - $T$) spin phase diagram for the CuGeO$_3$ sample obtained from the $M(H)$ and $\chi(T)$ measurement results.
 The diamond symbols represent data taken from a previous study\cite{Uchinokura:1995cu}. The dotted curves are guide for the eyes. 
 }
\label{}
\end{center}
\end{figure}

\thispagestyle{empty}
\begin{figure}
\begin{center}
\includegraphics[width=15 cm, bb= 0 0 223 105]{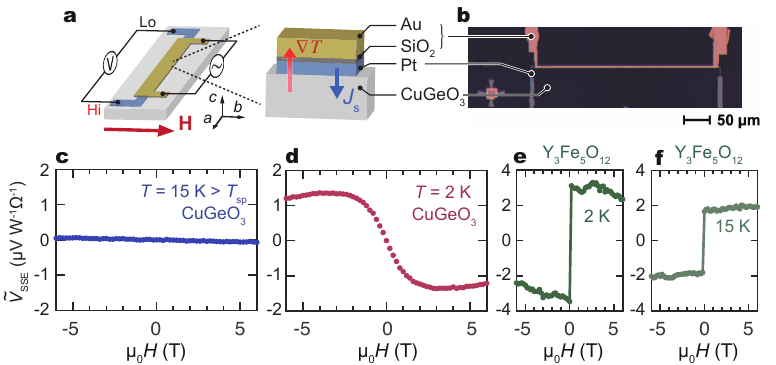}
\caption{ \textbf{SSE measurement results for CuGeO$_3$/Pt.}
\textbf{a,} Experimental setup for the SSE measurement. Temperature gradient ($\nabla T$) is applied by using an Au heater electrically insulated from the Pt layer by a SiO$_2$ film. An a.c. current is applied to the Au heater and SSE voltage in the Pt is measured with a lock-in amplifier as the second harmonic voltage $V_{2f}$.  
\textbf{b,} An optical micrograph for a on-chip SSE device made on the top of CuGeO$_3$.
\textbf{c} and \textbf{d,} The $H$ dependence of $\tilde{V}_{\mathrm{SSE}}$ (the SSE signal normalized to the heating power and the detector resistance) for the CuGeO$_3$/Pt sample at $15$ K and $2$ K, respectively.
\textbf{e}  and \textbf{f,}  $\tilde{V}_{\mathrm{SSE}}(H)$ measured for Y$_3$Fe$_5$O$_{12}$/Pt at 2 K and 15 K.
}
\label{fig2}
\end{center}
\end{figure}

\thispagestyle{empty}
\begin{figure}
\begin{center}
\includegraphics[width=13cm, bb= 0 0 219 216]{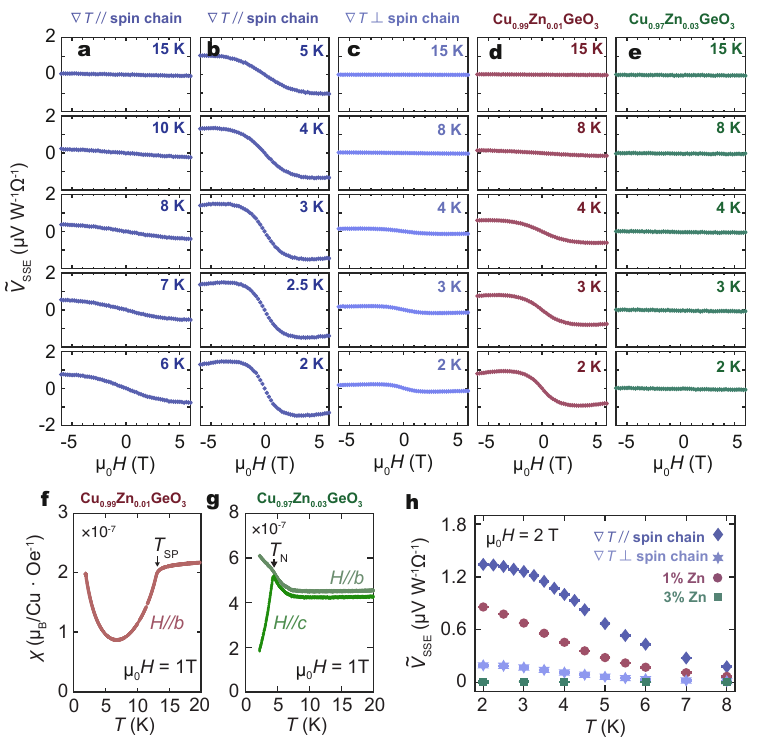}
\caption{\textbf{ $\mathbfit{T}$ dependence of $\mathbfit{V}_{\mathbf{SSE}}$.}
\textbf{a} and \textbf{b,} $H$ dependence of $\tilde{V}_{\mathrm{SSE}}$ at selected temperatures for CuGeO$_3$/Pt, with $\nabla T$ applied along the spin chain. 
\textbf{c,} $H$ dependence of $\tilde{V}_{\mathrm{SSE}}$ for CuGeO$_3$/Pt, with $\nabla T$ applied perpendicular to the spin chain. 
\textbf{d} and \textbf{e,} $H$ dependence of $\tilde{V}_{\mathrm{SSE}}$ at selected temperatures for Cu$_{0.99}$Zn$_{0.01}$GeO$_3$/Pt and Cu$_{0.97}$Zn$_{0.03}$GeO$_3$/Pt, respectively. In \textbf{a}-\textbf{e}, the SSE signal is normalized by the heating power and the detector resistance.
\textbf{f} and \textbf{g,} $T$ dependence of magnetic susceptibility for Cu$_{0.99}$Zn$_{0.01}$GeO$_3$ and Cu$_{0.97}$Zn$_{0.03}$GeO$_3$, respectively.
 \textbf{h,}  $T$ dependence of SSE signal in different samples at $\mu_0H = 2$ T. Error bars represent standard deviation.
}
\label{fig3}
\end{center}
\end{figure}

\thispagestyle{empty}
\begin{figure}
\begin{center}
\includegraphics[width=15cm, bb=0 0 177 73]{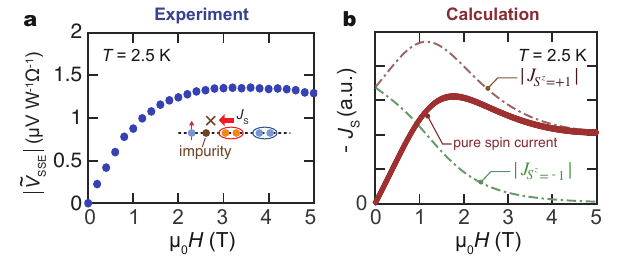}
\caption{\textbf{Magnetic field dependence of SSE voltage. } 
\textbf{a,} Magnetic field ($\mu_0H$) dependence of the magnitude of $|\tilde{V}_{\mathrm{SSE}}|$ at $2.5$ K.
\textbf{b,} Calculation results of the $\mu_0H$ dependence of the triplon spin current (see text and Supplementary Note G2 for details).}
\label{fig4}
\end{center}
\end{figure}



\section*{Supplementary material}

# Supplementary information for

# Triplon current generation in solids

Yao Chen, Masahiro Sato, Yifei Tang, Yuki Shiomi, Koichi Oyanagi, Takatsugu Masuda, Yusuke Nambu, Masaki Fujita, & Eiji Saitoh

**Supplementary Note A. Experimental details of spin-Seebeck effect measurements**

Single crystalline $CuGeO_3$ prepared by a floating zone method[1] was an elliptical cylinder with the height of $\sim 3$ cm (along the $a$-axis), the long axis of 7 mm (along the $c$-axis) and a short axis of 3 mm (along the $b$-axis). Crystal orientations are determined by using a Laue camara. Firstly, a 340 $\mu$m long, 1.5 $\mu$m wide and 5 nm thick Pt wire was fabricated on $CuGeO_3$ by an e-beam lithography and lift-off process. Subsequently, a $SiO_2$/Au wire with the length of 360 $\mu$m and the width of 7.5 $\mu$m was fabricated on the top of the Pt wire. The thickness of the $SiO_2$ and Au are 8 nm and 80 nm, respectively. Au and Pt layers are insulated by a $SiO_2$ layer.

In the SSE measurement, a sinusoidal current ($f = 13$ Hz) was applied to the Au layer with a current source (Keithley 6221, Tektronix, Inc.). The voltage signal showing up along the Pt wire was recorded by a lock-in amplifier (NF 5640, NF Corporation). All SSE measurements were performed in a PPMS (Physical Property Measurement System, Quantum Design, Inc.).

**Supplementary Note B. Estimation of $T_{\text{SP}}$ and $H_{\text{m}}$ of the CuGeO$_3$ sample**

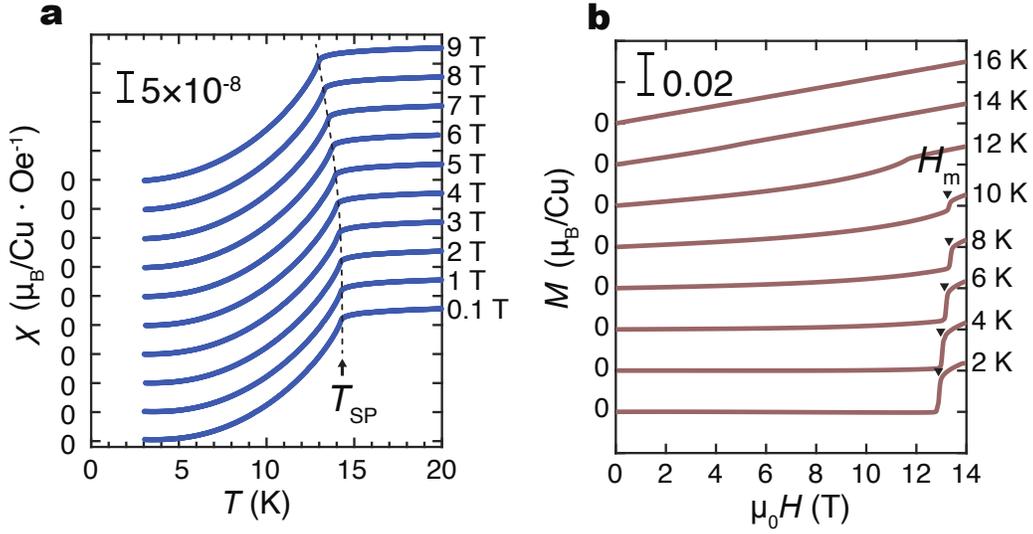

Supplementary Fig. 1. **Temperature ($T$) and magnetic field ($H$) dependence of the magnetization ($M$) of CuGeO$_3$.** **a,** $T$ dependence of magnetic susceptibility ($\chi = M/H$) at different $H$ values. **b,** $H$ dependence of $M$ at different $T$ values. Data are obtained with $H$ applied along the $b$-axis.

Magnetization ($M$) measurements were carried out using the VSM (vibrating sample magnetometer) option of PPMS. The $\chi(T) = M(T)/H$ results obtained under various $H$ values are shown in Supplementary Fig. 1a. Based on the results, we determined the spin-Peierls (SP) transition temperature $T_{\text{SP}}$ at each $H$. The transition field ($H_{\text{m}}$) from the SP phase to the magnetic phase at each $T$ was determined by measuring $M(H)$. A steep increase in $M(H)$ was detected at $\mu_0 H \sim 13$ T for $T < T_{\text{sp}}$, as shown in Supplementary Fig. 1b. $H_{\text{m}}$ is defined as the field where $dM(H)/dH$ is maximized.

**Supplementary Note C. Comparison between magnetization and SSE signal**

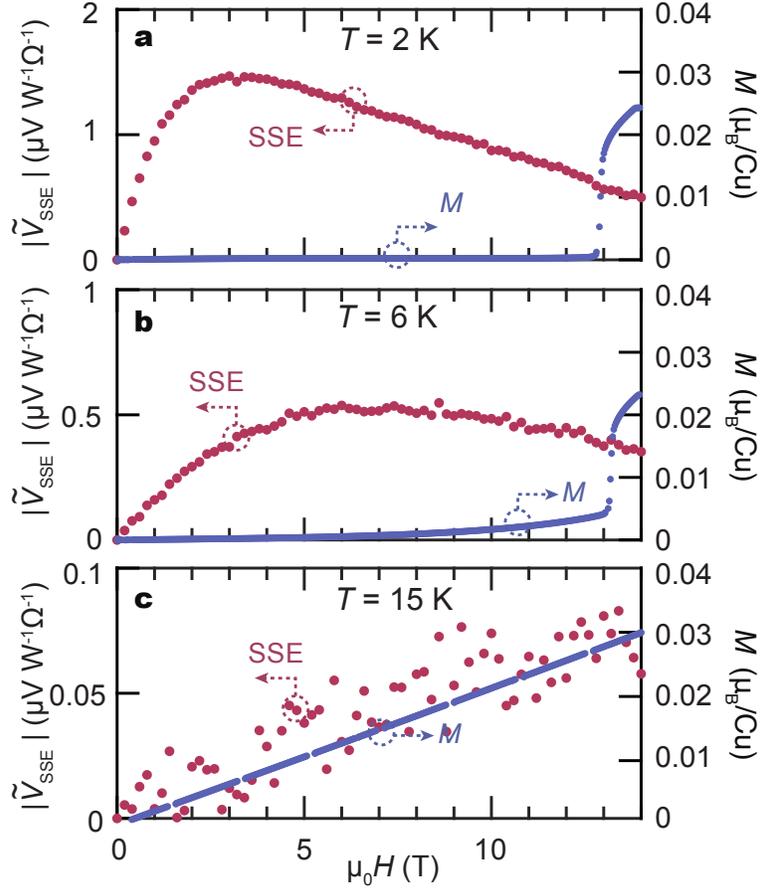

Supplementary Fig.2. **Comparison between magnetization $M$ and the SSE signal $|V_{\text{SSE}}|$. a-c,** at 2 K, 6 K and 15 K, respectively.

A comparison between $M(H)$ and SSE at different temperatures are shown in Supplementary Fig. 2. Unlike the case of the magnon SSE in most ferro/ferri-magnets, $V_{\text{SSE}}(H)$ in CuGeO$_3$ does not scale with $M(H)$. The magnetization of CuGeO$_3$ at $T < T_{\text{SP}}$ mainly consists of impurity spins and unpaired spins on thermally broken dimers. As shown in Supplementary Fig. 2a and 2b,

3

the SSE signal does not scale with $M$-$H$ at $T = 2$ K and 6 K. The $M$-$H$ in the low field range reflects the paramagnetic term from free (impurity) spins, and these free spins do not play a role in $V_{\mathrm{SSE}}(H)$ because they are well decoupled. At $T = 6$ K, the magnetic component from broken dimers overwhelms the paramagnetic component from impurity spins. Since the broken dimers are also little important for spin-current carriers in SSE, $V_{\mathrm{SSE}}$ and $M$ should show totally different magnetic field dependence.

For $T = 15$ K $> T_{\mathrm{SP}}$, the system is theoretically predicted to be a type of paramagnets: a Tomonaga-Luttinger (TL) spin liquid, where the equilibrium state is still a zero magnetizated state in the absence of fields and the spin excitation is a gapless spinon instead of triplon. For higher temperatures ($k_{\mathrm{B}}T > J$), the system experience a crossover from the TL spin liquid to a paramagnetic state. We also expect a spinon SSE[2] in the intermediate temperature range ($T_{\mathrm{SP}} < T \ll J/k_{\mathrm{B}}$). From the paper by Hirobe, et al (ref. [2]), the spinon SSE signal in TL spin liquids was found to be much smaller than the ferromagnetic SSE, linear with respect to the magnetic field. The SSE signal measured at 15 K $> T_{\mathrm{SP}}$ satisfies all these properties, and thus may be an indication of the spinon SSE.

Also, the anomaly in the SSE is hardly observed at $H_{\mathrm{m}}$. This is because the net spin current above $H_{\mathrm{m}}$ is still determined by low-energy $S = +1$ triplons and high-energy $S = -1$ triplons, similar to the case of $H < H_{\mathrm{m}}$ (see Supplementary Fig. 3 which illustrates the triplon bands above $H_{\mathrm{m}}$). As shown in Supplementary Fig. 3, for $H > H_{\mathrm{m}}$, the excitation energy of the $S = +1$ triplon is lower than the energy of the ground state, and the $S = +1$ triplon condenses. This is

4

called the triplon BEC (ref.[3]). The condensed triplon shows static magnetization but does not carry spin current.

The comparison between SSEs in an antiferromagnetic insulator and the SP magnet $CuGeO_3$ would be useful to deeply understand the feature of the triplon SSE. In the case of antiferromagnet/Pt system, the SSE signal changes abruptly near the spin-flop transition[4,5]. At the spin-flop transition, both the ground state and the magnon-band structure drastically change and a net magnetization suddenly appears because the spin-flop transition is of a first-order type. Due to this significant change of the magnon band, the SSE signal largely changes at the spin-flop transition. On the other hand, although the triplon BEC occurs for $H > H_\mathrm{m}$ of $CuGeO_3$, the band structure of the triplon excitation is very similar to that for $H < H_\mathrm{m}$. Therefore, there was no significant change in the triplon SSE signal around $H_\mathrm{m}$.

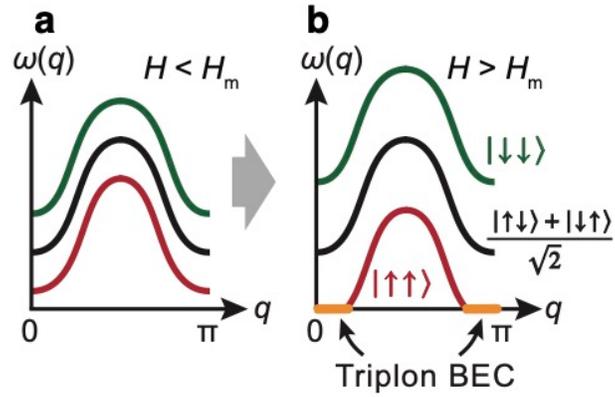

Supplementary Fig.3. **A schematic illustration of triplon BEC. a,** triplon band structure with magnetic field ($H$) less than the transition field ($H_\mathrm{m}$) from the SP phase to the magnetic phase. **b,** triplon band structure with $H > H_\mathrm{m}$. Green, black and red curves represent three triplet states $|\downarrow\downarrow\rangle$, $\left(|\uparrow\downarrow\rangle + |\downarrow\uparrow\rangle\right)/\sqrt{2}$, and $|\uparrow\uparrow\rangle$ respectively.

## Supplementary Note D. Angular and heater-power dependence of the spin-Seebeck signal

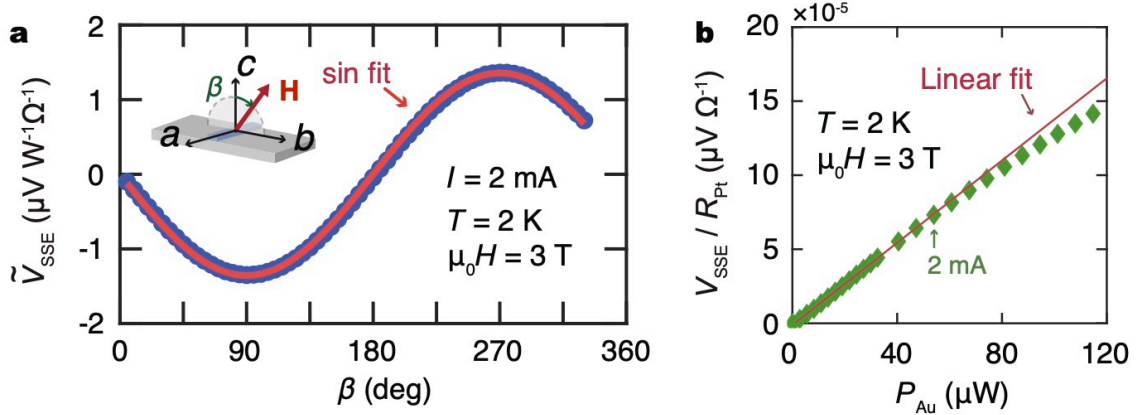

Supplementary Fig.4. **Angular and heating power dependence of the SSE signal. a,** Angular dependence of $\tilde{V}_{\mathrm{SSE}}$ at $\mu_0 H = 3$ T and $T = 2$ K, with a heating current ($I$) of 2 mA. The red curve indicates a sinusoidal fit. **b,** Heater-power dependence of $\tilde{V}_{\mathrm{SSE}}$ at $\mu_0 H = 3$ T and $T = 2$ K. The red line is a linear fit to the data for the heater currents ranging from 0 to 2 mA. Error bars represent 95% confidence interval of the sinusoidal fitted amplitude.

Supplementary Fig. 4a shows $\tilde{V}_{\mathrm{SSE}}(\beta)$ in CuGeO$_3$/Pt at $T = 2$ K and $\mu_0 H = 3$ T. Here, $H$ is always perpendicular to the Pt wire and the out-of-plane magnetic field angle $\beta$ is defined as shown in Supplementary Fig. 4a. $\tilde{V}_{\mathrm{SSE}}(\beta)$ is well fitted by a $\sin(\beta)$ function, which is consistent with the angular dependence of the inverse spin-Hall effect (ISHE) voltage: $V_{\mathrm{ISHE}} \propto [\mathbf{J}_\mathrm{s} \times \boldsymbol{\sigma}]_a \propto \sin(\beta)$. Here, $\mathbf{J}_\mathrm{s}$ and $\boldsymbol{\sigma}$ denote the spatial direction of the spin current and the spin-polarization vector of the spin current, respectively[6]. $[\mathbf{J}_\mathrm{s} \times \boldsymbol{\sigma}]_a$ is the $a$-axis component of $\mathbf{J}_\mathrm{s} \times \boldsymbol{\sigma}$.

The heater-power dependence of $V_{\mathrm{SSE}}$ is shown in Supplementary Fig. 4b. The amplitude

7

of $V_{\text{SSE}}$ at each heater-power is estimated by fitting $V_{\text{SSE}}(\beta)$ using a $\sin(\beta)$ curve at $\mu_0 H = 3$ T. When the current applied to the Au heater is less than 2 mA, the heater-power dependence of $V_{\text{SSE}}$ is well fitted by a linear function. For higher heater power, however, $V_{\text{SSE}}$ deviates from the linear dependence. This non-linear dependence was also observed in the magnon spin-Seebeck effect in $Y_3Fe_5O_{12}$/Pt systems[7]; the non-linearity is attributed to the strong non-equilibrium excitation of magnons in $Y_3Fe_5O_{12}$. Another possibility for the non-linearity in the $CuGeO_3$ case is temperature rise of the $CuGeO_3$ sample due to the Joule heating, since $V_{\text{SSE}}$ should decrease with increasing $T$. Therefore, to avoid the undesirable non-equilibrium/heating effects, we fixed the heater current at 1.5 mA for all the measurements shown in the main text.

**Supplementary Note E. Temperature and field dependence of the Pt and Au resistivity**

$V_{\text{SSE}}$ is proportional to the heating power of the Au heater. In the case of a constant current ($I_{\text{Au}}$), $V_{\text{SSE}}$ is proportional to the resistance of Au ($R_{\text{Au}}$). $V_{\text{SSE}}$ is also proportional to the resistance of Pt ($R_{\text{Pt}}$) for a constant spin current $J_s$ injection: $V_{\text{SSE}} \propto J_s \times R_{\text{Pt}}$. Thus, a change in $R_{\text{Pt}}$ and $R_{\text{Au}}$ with respect to $T$ and $H$ will cause extrinsic variations in $V_{\text{SSE}}$. Supplementary Figs. 5a and b show the $T$ dependence of $R$ of the Pt wire and Au heater, respectively. As $T$ is reduced, both $R_{\text{Pt}}$ and $R_{\text{Au}}$ decrease monotonically. As shown in the insets of Supplementary Figs. 5a and b, the variation of $R_{\text{Pt}}$ and $R_{\text{Au}}$ at low $T$ ($T < 20$ K) is less than 1 %. The $H$-dependence of $R_{\text{Pt}}$ and $R_{\text{Au}}$ at $T = 2$ K is shown in Supplementary Figs. 5c and S5d. $R_{\text{Pt}}$ and $R_{\text{Au}}$ change less than 1 % up to 6 T in both metals. According to the $R(T)$ and $R(H)$ results, the contribution of the resistance change to the observed SSE signal is negligibly small.

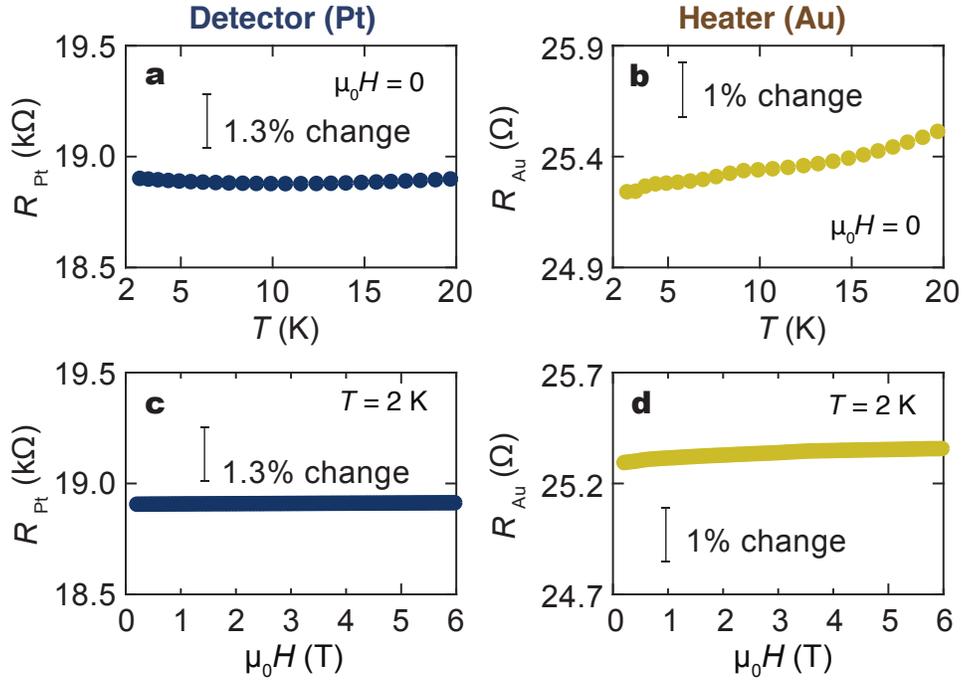

Supplementary Fig.5. **Temperature ($T$) and magnetic field ($H$) dependence of resistance ($R$) of detector and heater. a-b,** $T$ dependence of $R$ of Pt and Au. **c-d,** The magnetoresistance of Pt and Au at 2 K. The resistivity change for both metals up to 6 T is below 1 %.

As for the actual temperature at the interface, we can use the Pt detector layer as a thermometer (i.e. Pt resistance thermometer) to investigate the temperature change with and without the Au heating current (e.g. ref.[8]). When the heater current is applied to the Au heater, the resistance of the Pt detector changes, as shown in Supplementary Fig. 6a. Using $dR/dT$ and $\Delta R \equiv R(I_{Au} = 1.5\text{mA}) - R(I_{Au} = 0)$, we estimated the typical temperature difference across the device $\Delta T = \Delta R/(dR/dT)$ to be 0.2-0.5 K, as shown in Supplementary Fig. 6d. For $T > 5$ K, $\Delta T$ is as small as 0.25 K and 0.5 K even at $T = 2$ K. In the constant power setup, a lower value of

thermal conductivity $\kappa$ leads to a larger temperature difference and causes a larger observed SSE signal. In the case of CuGeO$_3$, because of the suppression of phonon transport, $\kappa$ monotonically decreases for $T < 5$ K (ref. [9]), which is consistent with our results.

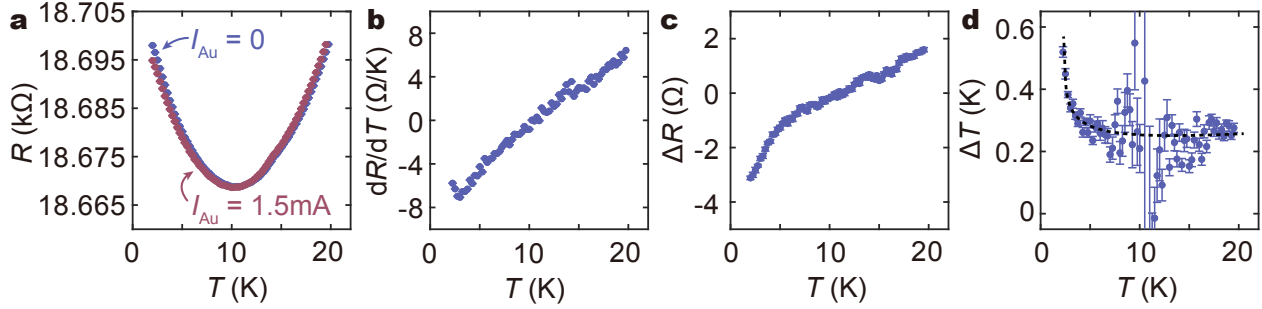

Supplementary Fig.6. **Temperature difference at the interface. a,** The temperature ($T$) dependence of the resistance ($R$) of Pt with and without the heating current ($I_\text{Au}$). **b-d,** The temperature dependence of $\text{d}R/\text{d}T$, $\Delta R$ and $\Delta T$. The dotted curve is a guide for the eyes. Error bars represent standard deviation.

**Supplementary Note F. Impurity density in Cu$_{0.99}$Zn$_{0.01}$GeO$_3$ and nominally pure CuGeO$_3$**

The density of impurities in CuGeO$_3$ can roughly be estimated from the $\chi(T)$ curve. In the SP phase, the magnetic susceptibility can be decomposed into three terms[10,11]:

$$\chi(T) = \chi_0 + \chi_\text{Para}(T) + \chi_\text{SP}(T) \tag{1}$$

where $\chi_0$, $\chi_\text{Para}$, and $\chi_\text{SP}$ stand for constant diamagnetic contribution, paramagnetic contribution from impurity-induced free spins (unpaired Cu spins) obeying $\chi_\text{Para} = C/(T - \Theta)$, and spin-Peierls contribution, respectively. For Cu$_{0.99}$Zn$_{0.01}$GeO$_3$ sample, $C = 5.1 \times 10^{-7}$ $\mu_\text{B}\text{Cu} \cdot \text{Oe}^{-1}$

10

were obtained by fitting the $\chi(T)$ data by using Supplementary Eq. (1) at a low $T$ range ($< 3$ K), assuming that the spin-Peierls contribution is zero ($\chi_{\text{SP}} = 0$)[11]. The fitted result of $\chi_{\text{Para}}$ is shown as a function of $T$ in Supplementary Fig. 7a. Assuming that all $S = 1/2$ $Cu^{2+}$ ions contribute to the paramagnetic term, $C_{\text{all}} = \frac{S(S+1)g^2\mu_B^2}{3k_B} = 7.06 \times 10^{-5} \mu_B \text{Cu} \cdot \text{Oe}^{-1}$ with $g = 2.1$ for $CuGeO_3$ (ref. [12]). By comparing $C_{\text{all}}$ with the fitted $C$ result, we conclude that the density of free spins is about $C/C_{\text{all}} \sim 0.73\%$ of all Cu atoms, which is very close to the $1\%$ Zn-doping. Preforming the same analysis on the $CuGeO_3$ sample (see Supplementary Fig. 7b), we obtain an impurity density level ($C/C_{\text{all}}$) of $0.02\%$ in the $CuGeO_3$ sample. Therefore, even in nominally pure samples, the effect of triplon scattering by impurities cannot be ignored.

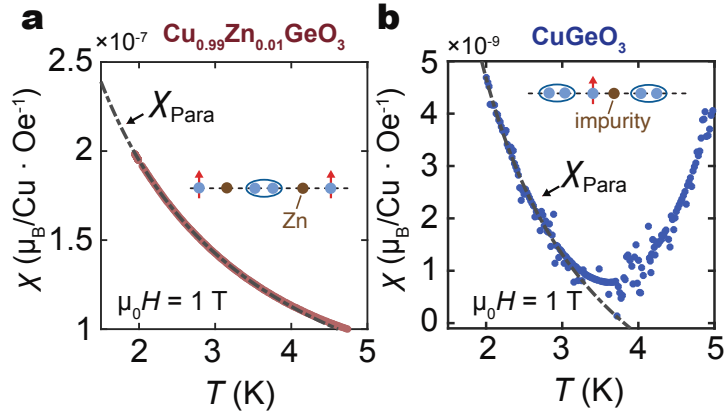

Supplementary Fig.7. **Magnetic susceptibility ($\chi$) at the low temperature ($T$) range. a** and **b,** temperature ($T$) dependence of magnetic susceptibility ($\chi$) for $Cu_{0.99}Zn_{0.01}GeO_3$ and $CuGeO_3$ at a low $T$ range. The impurity induced paramagnetic term $\chi_{\text{Para}}$ for each sample is estimated from the fitting to the Curie-Weiss law (black curves).

**Supplementary Note G. Theoretical analysis of SSE in CuGeO$_3$**

In this section, we discuss our theoretical analysis for SSE in CuGeO$_3$. First, we review a quantum sine-Gordon (SG) model as low-energy effective theory for CuGeO$_3$. The SG model describes triplon excitation in the spin-Peierls phase. Then, based on the results of the SG model, we compute a SSE spin current in CuGeO$_3$. To this end, we apply two microscopic methods: the Boltzmann equation and the tunnel spin-current theory based on Keldysh Green's function.

**Supplementary Note G1. Quantum sine-Gordon field theory as the low-energy effective model for CuGeO$_3$**

Here, we explain low-energy effective theory for CuGeO$_3$. The model Hamiltonian for each antiferromagnetic spin chain in CuGeO$_3$ is

$$H_{\rm SP} = \sum_j J_1(1+\delta(-1)^j)\mathbf{S}_j \cdot \mathbf{S}_{j+1} + J_2 \mathbf{S}_j \cdot \mathbf{S}_{j+2} - B\sum_j S_j^z. \tag{2}$$

where $\mathbf{S}_j$ is the electron spin on the $j$-th site, and $B \equiv g\mu_B\mu_0 H$ with the $g$ factor $g = 2.1$ (ref. [12]), Bohr magneton $\mu_B$ and an external magnetic flux density $\mu_0 H$. For CuGeO$_3$, the exchange coupling constants of nearest neighboring spins and next nearest neighboring spins are respectively estimated to be $J_1 \simeq 120.6$ (ref. K[13]) and $J_2 \simeq 0.36 J_1$ (ref. [14]). A finite dimerization parameter, $\delta$, emerges in the spin-Peierls phase due to the lattice deformation. First, we focus on the zero-field case. In the small-$J_2$ case of $\alpha \equiv J_2/J_1 < 0.241 \equiv \alpha_c$ (ref. [15]), the spin chain belongs to the universality class of a Tomonaga-Luttinger (TL) liquid, i.e., the ground state is a non-magnetic spin liquid and the low-energy excitations are described by gapless spinons. For $\alpha > \alpha_c$, the ground state is spontaneously dimerized and an excitation gap opens. A dimerization induces

a lattice distortion through the spin-phonon coupling, and, as a result, $\delta$ becomes finite. In the dimerized (i.e., spin-Peierls) phase with a finite $\delta$, the lowest excitation branch is given by the triply degenerated $S = 1$ mode, namely, triplons. The low-energy effective Hamiltonian [16–18] is given by

$$H_\text{eff} = \int dx \ \frac{v}{2\pi} \left[\frac{1}{K}(\partial_x\phi)^2 + K(\partial_x\theta)^2\right] + \frac{J_1\delta d}{a_0} \sin(\sqrt{4\pi}\phi) + \frac{\lambda}{a_0} \cos(\sqrt{16\pi}\phi). \quad (3)$$

Where $\phi$ and $\theta$ are a pair of dual boson fields. $v$ and $K$ are, respectively, the spinon velocity and the TL-liquid parameter, and $a_0$ is the lattice constant. The spin-rotation SU(2) symmetry makes the TL-liquid parameter fixed to be $K = 1/2$. The quadratic-boson (i.e., free-boson) part corresponds to the TL-liquid Hamiltonian. The sine term stems from the bond alternation $\delta$ and the non-universal constant $d(> 0)$ can be numerically determined [17,19] while the cosine term $\cos(\sqrt{16\pi}\phi)$ exists even in the uniform spin chain without $\delta$. The sine term with scaling dimension $K$ is more relevant than the cosine term with dimension $4K$. If $\alpha$ is close enough to $\alpha_c$, the coupling constant $\lambda \propto \alpha - \alpha_c$. In the uniform case of $\delta = 0$, the $\lambda$ term is marginally irrelevant (relevant) for $\lambda < 0$ ($\lambda > 0$), which corresponds to a gapless TL liquid (a spontaneously dimerized ground state with a spin gap). For a SP phase with $\delta \neq 0$, a small $\lambda$ term can be negligible in the sense of renormalization group, and the low-energy Hamiltonian is given by a sum of the TL-liquid part and a single cosine term $\cos(\sqrt{4\pi}\phi)$. This is called quantum sine-Gordon (SG) model and is an integrable quantum field theory [16,18,20,21]. Since $\alpha$ is close to $\alpha_c$ in CuGeO$_3$, we can expect that its low-energy physics is described by a SG model. The SG model is known to well describe the low-energy physics of the spin-Peierls Hamiltonian with high accuracy [16–19], and the model enables us to compute various physical quantities in a quantitative level [22,23].

13

The sine term is relevant in the SU(2) case we focus on, and therefore the field $\phi$ is locked at the energy minimum of the sine potential in the ground state, i.e., $\langle\phi\rangle = \sqrt{\pi}(n - \frac{1}{4})$ for $\delta > 0$ and $\langle\phi\rangle = \sqrt{\pi}(n + \frac{1}{4})$ for $\delta < 0$ ($n$: integer). The elementary excitations in the SG model are soliton, antisoliton, and breathers (bound state of soliton and antisoliton) [17,20,21]. The soliton and antisoliton are the domain-wall excitations connecting two neighboring ground states: For instance, a soliton texture located around $x = x_0$ is given by $\langle\phi\rangle = \sqrt{\pi}(n - \frac{1}{4})$ for $x \ll x_0$ and $\langle\phi\rangle = \sqrt{\pi}(n + 1 - \frac{1}{4})$ for $x \gg x_0$. The mass gap of these excitations depends on the parameters $v$, $K$, and $J_1 \delta d/a_0$ if we neglect the marginal cosine term $\cos(\sqrt{16\pi}\phi)$. In the SU(2) case, soliton, antisoliton, and first breather have the same mass $m_1 \sim \delta^{2/3}$. Second breather's mass $m_2$ is given by $m_2 = \sqrt{2} m_1$. In the language of spin-Peierls chains, the triply-degenerated excitations of soliton, antisoliton, and first breather correspond to spin-1 triplons around the wave number $k = \pi$, while the second breather is spin-singlet excitation around $k = \pi$. The energy bands of triplons, i.e., soliton ($S_z = 1$), first breather ($S_z = 0$), and anti-soliton ($S_z = -1$), are respectively given by

$$\epsilon_S(k_\pi) = \sqrt{m_1^2 + v^2 k_\pi^2} - B, \quad \text{for } S_z = 1; \tag{4}$$

$$\epsilon_B(k_\pi) = \sqrt{m_1^2 + v^2 k_\pi^2}, \quad \text{for } S_z = 0; \tag{5}$$

$$\epsilon_{AS}(k_\pi) = \sqrt{m_1^2 + v^2 k_\pi^2} + B, \quad \text{for } S_z = -1. \tag{6}$$

where $k_\pi = k - \pi$. The $B$ dependence of $\epsilon_{AS}$ and $\epsilon_S$ reflects the Zeeman splitting.

Finally, we mention values of the parameters in the SG model for $CuGeO_3$. We have $K = 1/2$ in the SU(2) model, and the spin gap (triplon gap) $m_1 \simeq 2.36$ meV ($\approx 27.4$ K) is estimated from an ESR experiment[12] at a low temperature. The velocity $v$ may be approximated by that of $J_1$-

$J_2^c$ spin chains since $\alpha$ of CuGeO$_3$ is closed to $\alpha_c$. The velocity has been numerically estimated as $v = 1.174 J_1 a_0$ (refs. [17,24]). Combining this value of $v$ and the exact result of the SG model[17,20], we can compute the soliton mass as $m_1/J_1 \simeq 2.13\delta^{2/3}$. This relation, $J_1 \simeq 120.6$ K, and $m_1 \simeq 27.4$ K lead to $\delta \simeq 0.034$ in CuGeO$_3$ at a sufficiently low temperature. If temperature $T$ grows up to the same order as the lowest gap $\epsilon_S(0) = m_1 - B$, the fluctuation of $\delta$ increases mainly due to the effect of phonons, and as a result, the description of the SG model gradually becomes invalid.

**Supplementary Note G2. Approach based on Boltzmann equation**

Using the above result, let us compute the thermal spin current in SSE for CuGeO$_3$. The SG model predicts that when magnetic field $B$ is increased in the SP phase, the difference of the soliton density and anti-soliton one (i.e., spin current carrier) monotonically increases at a fixed temperature due to the Zeeman splitting. Nevertheless, the experimental result shows a non-monotonic $B$ dependence of the SSE voltage. It indicates the importance of scattering processes among triplons, impurities, phonons, etc.

If we consider a triplon as a wave packet located at $r$ with the wave vector $\mathbf{k}$, the triplon distribution function $f_\eta(\mathbf{k}, \mathbf{r}, t)$ follows the Boltzmann equation,

$$\frac{\partial}{\partial t} f_\eta + \mathbf{v}_\eta(\mathbf{k}) \cdot \nabla_\mathbf{r} f_\eta + \frac{d\mathbf{k}}{dt} \cdot \nabla_\mathbf{k} f_\eta = \frac{\partial}{\partial t} f_\eta |_{\text{scattering}}, \qquad (7)$$

where the index $\eta$ corresponds to soliton ($S_z = -1$) and anti-soliton ($S_z = 1$). The right hand side is the scattering rate of triplons. The group velocity of trilpon is $\mathbf{v}_\eta(\mathbf{k}) = \frac{1}{\hbar} \frac{\partial \epsilon_\eta(\mathbf{k})}{\partial \mathbf{k}}$. Since breathers are spin-singlet quasi particles and do not contribute to spin current. Therefore, it is enough to consider soliton and anti-soliton for computing the spin current.

The real compound CuGeO$_3$ has a weak but finite three dimensionality, but we expect that the triplon distribution along the spin-chain ($x$) direction is essential for the SSE in CuGeO$_3$. Therefore, we will replace **k** with the wave number $k$ along the $x$ direction hereafter. In the SSE measurement, a temperature gradient is applied along the $x$ axis to the sample. If the system length $L_x$ is sufficiently large, we may approximate the gradient as the constant $\partial T/\partial x = -\frac{T_{\text{high}} - T_{\text{low}}}{L_x} \equiv \Delta T_x$. For a small $\Delta T_x$, the system approaches to a non-equilibrium steady state following local equilibrium, and the distribution function is approximated as

$$f_\eta = f_\eta^{(0)}(k) + g_\eta(k, x, t), \tag{8}$$

where $f_\eta^{(0)}(k)$ is the equilibrium distribution function and $g_\eta(k, x, t)$ is the deviation from the equilibrium. We assume that $f_\eta^{(0)}$ can be described by a Bose distribution function,

$$f_\eta^{(0)}(k) = \frac{1}{e^{\beta(\epsilon_\eta(k) - \mu)} - 1}. \tag{9}$$

Triplons are not equal to bosons or fermions, but the approximation of Supplementary Eq. (9) is expected to work well if triplon density is low enough. The non-equilibrium steady state satisfies $\partial f_\eta/\partial t = dk/dt = 0$. For the scattering term, we adapt relaxation time ($\tau_k$) approximation [25]:

$$\frac{\partial}{\partial t} f_\eta \Big|_{\text{scattering}} = -\frac{f_\eta(k, x, t) - f_\eta^{(0)}(k)}{\tau_k} = -\frac{g_\eta(k, x, t)}{\tau_k}. \tag{10}$$

Under the assumption that $\partial T/\partial x$ and $g$ are small, we may omit the $\partial g/\partial T$ term in Supplementary Eq. (7) and therefore it is simplified as

$$g_\eta(k) = \tau_k v_\eta(k) k_B \Delta_x T \frac{\partial}{\partial (k_B T)} f_\eta^{(0)}(k) \tag{11}$$

From these discussions, the total spin current with $S^z$ polarization along the $x$ direction is

$$\begin{aligned} J_s &= \sum_{S_z=1,-1} \sum_k \hbar S_z v_{S_z}(k) f_{S_z}(k) \\ &= \sum_{S_z} \sum_k \hbar S_z v_{S_z}(k) g_{S_z}(k) \\ &= \sum_{S_z} \sum_k \hbar S_z v_{S_z}(k)^2 \tau_k k_B \Delta_x T \frac{\epsilon_{S_z}(k)}{(k_B T)^2} \frac{e^{\beta \epsilon_{S_z}(k)}}{(e^{\beta \epsilon_{S_z}(k)} - 1)^2} \\ &= \sum_{S_z} \int_{\text{B.Z.}} \frac{dk}{2\pi} \hbar S_z v_{S_z}(k)^2 \tau_k k_B \Delta_x T \frac{\epsilon_{S_z}(k)}{(k_B T)^2} \frac{e^{\beta \epsilon_{S_z}(k)}}{(e^{\beta \epsilon_{S_z}(k)} - 1)^2}. \end{aligned} \qquad (12)$$

This is Equation (1) in the main text.

To numerically calculate the $H$ dependence of $J_s$, it is necessary to determine the relaxation time $\tau_k$ based on appropriate impurity scattering mechanisms. If we focus on the elastic scattering between a triplon and an impurity, $\tau_k$ is estimated as [25]

$$\tau_k(\epsilon)^{-1} \sim n_{\text{imp}} D(\epsilon) V^2, \qquad (13)$$

where $n_{\text{imp}}$, $D(\epsilon)$ and $V$ are impurity density, the density of states (DOS) of the triplons and impurity potential, respectively. In CuGeO$_3$, $J_c \sim 120$, $J_b \sim 0.1 J_c$ and $J_a \sim -0.01 J_c$ (ref. [13]), and the two-dimensionality is strong compare to other one-dimensional systems. Therefore, We may approximate $D(\epsilon)$ at the bottom of the band as a constant. Consider the elastic scattering from both nonmagnetic and magnetic impurities, the total relaxation time can be approximated as

$$\tau_{k,\text{total}}^{-1} = \tau_{k,\text{non-mag}}^{-1} + \tau_{k,\text{mag}}^{-1}, \qquad (14)$$

if the impurity densities are small enough. Combining these theoretical arguments and experimental results on the doped CuGeO$_3$, we assume that the relaxation times from nonmagnetic and

magnetic impurities is given by

$$\tau^{-1}_{k,\text{non}-\text{mag}} = n_{\text{imp}} C_0 V_0^2; \qquad (15)$$

$$\tau^{-1}_{k,\text{mag}} = \left(n_{\text{imp}} B_S(T, H)\right) \cdot C_{\text{mag}} \cdot \left(V_{\text{mag}} B_S(T, H)\right)^2. \qquad (16)$$

$C_0$ and $C_{\text{mag}}$ are constants and $B_S(T, H)$ is the Brillouin function with $S = 1/2$. The introduction of $B_S$ is reasonable because it is expected that magnetically-polarized impurities more strongly refute triplons compared to weakly polarized ones. The numerical result of $J_s(H)$ at $T = 2.4$ K is shown in the main text as Fig. 4, in which $C_{\text{mag}} V_{\text{mag}} \gg C_0 V_0$ and the $T$ dependent triplon gap $m_1(T)$ measured in the neutron scattering are used. Due to the Zeeman-splitting induced density difference between soliton and antisoliton, the sign of spin current is opposite to that in ferromagnetic insulators. We verify that the broad peak structure of $J_s$ as a function of $H$ is stable against small changes of $\tau^{-1}_{k,\text{mag}}$. Moreover, the broad peak still survives even if the equilibrium distribution $f^{(0)}_\eta$ is changed into the fermion one in the regime of a low triplon density.

**Supplementary Note G3. Approach based on the tunnel spin current**

Another microscopic approach for SSE is the formula of tunnel spin current on the interface from the magnet (CuGeO$_3$) to the metal (Pt)[2,26–28]. An advantage of this approach is that it can be applied to a broad class of magnets even with non-magnon type excitations such as spinons, triplons, magnon pairs, etc. However, it cannot take into account the effects of scattering among magnetic excitations during a flow of spin currents.

In this approach, we focus on the interface between CuGeO$_3$ and Pt. We assume that the dominant interaction at the interface is Heisenberg type exchange interaction between localized

spin of $CuGeO_3$ and conducting-electron spin of Pt. Such an exchange interaction has been used to describe interfacial spin transfer between an insulator magnet and a metal[2,26–28]. The interface Hamiltonian is given by

$$H_{\text{int}} = J_{\text{int}} \sum_{\mathbf{r}\in\text{interface}} \mathbf{S_r} \cdot \mathbf{s_r}, \tag{17}$$

where $\mathbf{S_r}$ and $\mathbf{s_r}$ respectively denote the localized spin and the conducting-electron spin at site $\mathbf{r}$ on the interaface. If we perturbatively treat this interface interaction using Keldysh Green's function method [27,29], the tunneling spin current is calculated as

$$I_s \propto -J_{\text{int}}^2 \int_{-\infty}^{\infty} d\omega \text{Im} X_R^{-+}(\omega) \text{Im} \chi_R^{-+}(\omega) \left[\coth\left(\frac{\omega}{2k_B T_{\text{metal}}}\right) - \coth\left(\frac{\omega}{2k_B T_{\text{magnet}}}\right)\right] \tag{18}$$

up to the leading order of the interface coupling constant $J_{\text{int}}$. Here, $\omega$, $T_{\text{metal}}$ and $T_{\text{magnet}}$ are respectively frequency, the spatially averaged value of temperature in the metal (Pt), and that in the $CuGeO_3$. $X_R^{-+}(\omega)$ and $\chi_R^{-+}(\omega)$ are respectively the retarded part of the local dynamical susceptibility of the $CuGeO_3$ at $T = T_{\text{magnet}}$ and that of the Pt at $T = T_{\text{metal}}$. The indices $^{-+}$ denotes the transverse spins $S^{\pm}$. $-\text{Im} X_R^{-+}(\omega)$ with $\omega > 0$ may be viewed as the DOS of $S_z = +1$ magnetic excitations, while $\text{Im} X_R^{-+}(\omega)$ with $\omega < 0$ as that of $S_z = -1$ magnetic excitations[2]. On the other hand, the magnetic energy scale in $CuGeO_3$ is much smaller than that of the kinetic energy of electrons in the metal (Pt), and therefore the susceptibility of the metal can be approximated as[30,31]

$$\text{Im}\chi_R^{-+}(\omega) \simeq \hbar\omega D(\epsilon_F)^2 + \cdots, \tag{19}$$

where $D(\epsilon_F)$ is the DOS of conduction electrons at Fermi surface $\epsilon = \epsilon_F$ in the metal. Namely, $\text{Im}\chi_R^{-+}(\omega)$ is an odd function of $\omega$. Therefore, the sign of the tunnel spin current reflects the difference between the weights of $S_z = -1$ and $S_z = +1$ modes. If the temperature difference

$\Delta T = (T_{\text{magnet}} - T_{\text{metal}})/2$ is sufficiently small, the $T$-dependent factor in Supplementary Eq. (18) is approximated by

$$\coth\left(\frac{\omega}{2k_B T_{\text{metal}}}\right) - \coth\left(\frac{\omega}{2k_B T_{\text{magnet}}}\right) \simeq -\frac{\omega}{(k_B T_{\text{ave}})^2}\frac{1}{\sinh^2\left(\frac{\omega}{2k_B T_{\text{ave}}}\right)}\Delta T + \cdots, \quad (20)$$

where $T_{\text{ave}}$ is the averaged temperature $(T_{\text{magnet}} + T_{\text{metal}})/2$. Substituting Supplementary Eq. (19) and (20) into Supplementary Eq. (18), we arrive at the simplified formula

$$I_s \propto \pi J_{\text{int}}^2 D(\epsilon_F)^2 \Delta T \frac{1}{(k_B T)^2} \int_{-\infty}^{\infty} d\omega \text{Im} X_R^{-+}(\omega) \frac{\omega^2}{\sinh^2\left(\frac{\omega}{2k_B T}\right)}, \quad (21)$$

where for simplicity we have replaced $T_{\text{ave}}$ with $T$.

Here, we emphasise that the formula of this tunnel spin current has succeeded in well explaining several SSEs in standard ferro(i)magnets[26], an one-dimensional spin liquid[2], a spin-nematic magnet[28], and a compensated ferrimagnet[32]. We also note two points about the formula of Supplementary Eq. (18). The first thing is that triplons in the magnetic insulator $CuGeO_3$ cannot be injected to the metal Pt, while the spin of triplons is transferred to that of conducting electrons in Pt through the interface exchange $J_{\text{int}}$. Namely, spin angular momentum can tunnel from $CuGeO_3$ to Pt and vice versa. This is in contrast with a tunnel charge current in bilayer systems consisting of two metals. The second is about the possibility of the $T$ and $H$ dependences of $J_{\text{int}}$. The energy scale of chemical bonds at the interface is usually higher than those of the ranges of $T$ and $H$ (0 - 15 K and 0 - 14 T). Therefore, we have here assumed that $J_{\text{int}}$ is independent of $T$ and $H$.

The remaining task is to compute the susceptibility for a spin-Peierls chain $\text{Im} X_R^{-+}(\omega)$. If $CuGeO_3$ in the spin-Peierls phase is described by the SG model, we can use so-called the form-

20

factor method [21]. Following it, we calculate the dynamical susceptibility in $(k, \omega)$ space as

$$X_R^{-+}(k_\pi, \omega) \simeq \frac{-Z}{\omega + \epsilon_{\text{AS}}(k_\pi) + i\gamma} + \frac{Z}{\omega - \epsilon_{\text{S}}(k_\pi) + i\gamma} + \cdots, \qquad (22)$$

around $k = \pi$. Here the first and second terms are respectively the contribution of antisoliton ($S_z = -1$) and soliton ($S_z = 1$), and $\gamma \to +0$ is the infinitesimal factor. The renormalization factor $Z$ generally depends on $k$ and $\omega$, but it can be viewed as a constant if the system is ideally described by the integrable SG model. The local susceptibility $X_R^{-+}(\omega)$ is related to $X_R^{-+}(k_\pi, \omega)$ as $X_R^{-+}(\omega) = N^{-1} \sum_{k_\pi} X_R^{-+}(k_\pi, \omega)$ with $N$ being the total site number. As we mentioned, in finite temperature case, the neutron-scattering experiments showed that the mass gaps of soliton and antisoliton decrease and their life time becomes shorter. This behavior cannot be reproduced within the SG model, but such $T$ dependence can be treated by replacing $\gamma$ and $m_1$ of the bands $\epsilon_{\text{S,AS}}$ with the line width $\Gamma(T)$ and the $T$-dependent mass $m_1(T)$, respectively. Namely, we may describe the finite-temperature susceptibility $X_R^{-+}(k_\pi, \omega)$ as

$$X_R^{-+}(k_\pi, \omega) \simeq \frac{-Z}{\omega + \epsilon_{\text{AS}}(k_\pi, T) + i\Gamma(T)} + \frac{Z}{\omega - \epsilon_{\text{S}}(k_\pi, T) + i\Gamma(T)} + \cdots, \qquad (23)$$

where the bands $\epsilon_{\text{S,AS}}(k_\pi, T)$ are defined as $\epsilon_{\text{S}}(k_\pi, T) = \sqrt{m_1(T)^2 + v^2 k_\pi^2} - B$ and $\epsilon_{\text{AS}}(k_\pi, T) = \sqrt{m_1(T)^2 + v^2 k_\pi^2} + B$.

Using Supplementary Eq. (21) and Supplementary Eq. (23), we can calculate the $H$ and $T$ dependence of the tunnel spin current in a low-$T$ and low-$H$ regime. The calculated result shows that a spin current linearly emerges as the field $B$ increases and its sign is opposite to that of SSE in ferromagnetic insulators. These results in the small-$B$ regime are consistent with those from the Boltzmann equation.

**Supplementary Note G4. SSE in a moderate temperature regime**

To compute the $T$-dependence of $J_s$ in a relatively high temperature range (5 K - 10 K), we have to include the $T$-dependence of triplon bands and scattering processes in the Boltzmann equation approach. To this end, we need information on the $T$-dependence of the excitation gap of triplons $m_1(T)$ and the lifetime of triplons $1/\tau_k(T)$. By fitting neutron scattering results of spin-gap $m_1(T)$ (ref. [33]) with $m_1(T) = m_1(0)\left(1 - T/T_{\rm SP}\right)^\alpha$, we obtain $m_1(0) \sim 2.05$ meV and $\alpha \simeq 0.12$. In the numerical calculations, $m_1$ in Supplementary Eq. (4)-(6) is replaced by the fitted function of $m_1(T)$. The $T$-dependence of the lifetime of triplons is obtained as the peak width $\Gamma(T)$ of triplon excitation from inelastic neutron scattering experiments[33,34]. Here, the explicit $T$-dependence of $1/\tau_k$ is set as a new term $1/\tau_k(T)$, and the total relaxation time is

$$\tau_{k,\text{total}}^{-1} = \tau_{k,\text{non-mag}}^{-1} + \tau_{k,\text{mag}}^{-1} + \tau_k(T)^{-1}. \tag{24}$$

$1/\tau_{k,\text{non-mag}}$ is independent of $T$ and $1/\tau_{k,\text{mag}}$ only weakly depends on $T$ via the Brillouin function.

The experimental results of triplon peak width $\Gamma(T)$ obtained from Ref. [34,35] are well fitted by a power law $\Gamma(T) = \Gamma(0)T^4$, where $\Gamma(0)$ is the fitting parameter. The scattering with thermally excited phonons and other triplons may be two origins of the $T$-dependence of $\Gamma$ but the detail of microscopic mechanism is not clear. In this work, we treat these $T$-dependent scattering processes phenomenologically. Specifically, we set $\tau_k(T)$ as

$$\tau_k(T)^{-1} = C_{\text{th}} V_{\text{th}}^2 \Gamma(0) \left(\frac{T}{T_{\rm SP}}\right)^4. \tag{25}$$

Where $C_{\text{th}}$ and $V_{\text{th}}$ are $T$-independent constants.

Calculated result of the $T$-dependent SSE is shown in Supplementary Fig. 8. We have a

qualitative agreement between experiment and theory in the moderate temperature range (approximately 5 K - 10 K).

Finally, we emphasize that the result of the temperature dependence is justified only in the moderate temperature range. The reason why our result is less reliable in the high-$T$ range ($T \to T_{\mathrm{SP}}$) is as follows: our theory is based on the Sine-Gordon model, which can well describe the low-energy physics of the SP phase. When $T$ or $H$ increases, the dimerized ground state is partially broken down and the SG model gradually becomes less reliable. For $T \to T_{\mathrm{SP}}$, the critical nature of the SP transition becomes relevant, and the triplon picture is no longer applicable. This is also why theoretical calculations in Supplementary Notes G1-G3 focused only on the low temperatures and low magnetic fields, in which the singlet ground state and triplon excitation are well established. On the other hand, the relation of $\Gamma(T) \propto T^4$ is not experimentally supported in the low-$T$ range. Thus, Supplementary Fig. 8 shows the temperature range from 5 K to 11 K.

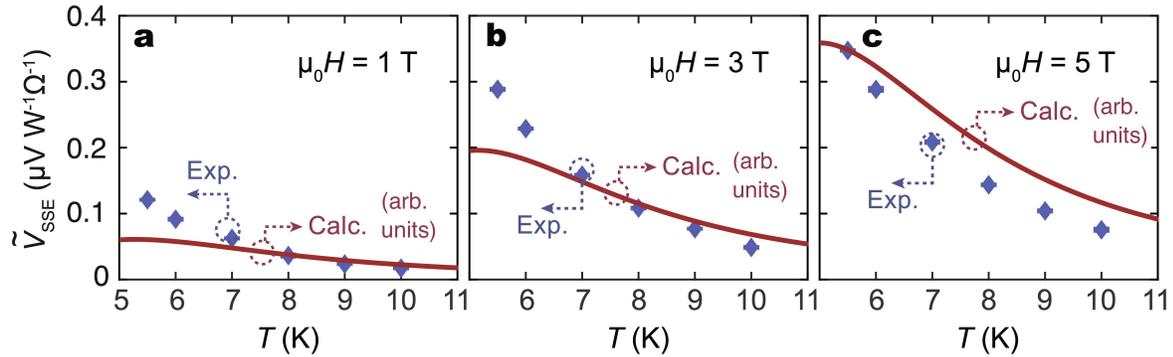

Supplementary Fig.8. **Comparison with theory. a-c** Comparison between calculation results of the $T$ dependence of the triplon spin current and the observed SSE signal at $\mu_0 H = 1, 3, 5$ T, respectively. Error bars represent standard deviation.




\newpage
\begin{thebibliography}{10}
\expandafter\ifx\csname url\endcsname\relax
  \def\url#1{\texttt{#1}}\fi
\expandafter\ifx\csname urlprefix\endcsname\relax\def\urlprefix{URL }\fi
\providecommand{\bibinfo}[2]{#2}
\providecommand{\eprint}[2][]{\url{#2}}

\bibitem{Uchida:2008cc}
\bibinfo{author}{Uchida, K.} \emph{et~al.}
\newblock \bibinfo{title}{{Observation of the spin Seebeck effect}}.
\newblock \emph{\bibinfo{journal}{Nature}} \textbf{\bibinfo{volume}{455}},
  \bibinfo{pages}{778--781} (\bibinfo{year}{2008}).

\bibitem{Takahashi:2010ju}
\bibinfo{author}{Takahashi, S.}, \bibinfo{author}{Saitoh, E.} \&
  \bibinfo{author}{Maekawa, S.}
\newblock \bibinfo{title}{{Spin current through a
  normal-metal/insulating-ferromagnet junction}}.
\newblock \emph{\bibinfo{journal}{Journal of Physics: Conference Series}}
  \textbf{\bibinfo{volume}{200}}, \bibinfo{pages}{062030}
  (\bibinfo{year}{2010}).

\bibitem{Saitoh:2006kk}
\bibinfo{author}{Saitoh, E.}, \bibinfo{author}{Ueda, M.},
  \bibinfo{author}{Miyajima, H.} \& \bibinfo{author}{Tatara, G.}
\newblock \bibinfo{title}{{Conversion of spin current into charge current at
  room temperature: Inverse spin-Hall effect}}.
\newblock \emph{\bibinfo{journal}{Applied Physics Letters}}
  \textbf{\bibinfo{volume}{88}}, \bibinfo{pages}{182509}
  (\bibinfo{year}{2006}).

\bibitem{Valenzuela:2006cs}
\bibinfo{author}{Valenzuela, S.~O.} \& \bibinfo{author}{Tinkham, M.}
\newblock \bibinfo{title}{{Direct electronic measurement of the spin Hall
  effect}}.
\newblock \emph{\bibinfo{journal}{Nature}} \textbf{\bibinfo{volume}{442}},
  \bibinfo{pages}{176--179} (\bibinfo{year}{2006}).

\bibitem{Ito:2019bq}
\bibinfo{author}{Ito, N.} \emph{et~al.}
\newblock \bibinfo{title}{{Spin Seebeck effect in the layered ferromagnetic
  insulators CrSiTe$_3$ and CrGeTe$_3$}}.
\newblock \emph{\bibinfo{journal}{Physical Review B}}
  \textbf{\bibinfo{volume}{100}}, \bibinfo{pages}{060402}
  (\bibinfo{year}{2019}).

\bibitem{2015NatPh..11.1022C}
\bibinfo{author}{Cornelissen, L.~J.}, \bibinfo{author}{Liu, J.},
  \bibinfo{author}{Duine, R.~A.}, \bibinfo{author}{Youssef, J.~B.} \&
  \bibinfo{author}{van Wees, B.~J.}
\newblock \bibinfo{title}{{Long-distance transport of magnon spin information
  in a magnetic insulator at room temperature}}.
\newblock \emph{\bibinfo{journal}{Nature Physics}}
  \textbf{\bibinfo{volume}{11}}, \bibinfo{pages}{1022--1026}
  (\bibinfo{year}{2015}).

\bibitem{Seki:2015esa}
\bibinfo{author}{Seki, S.} \emph{et~al.}
\newblock \bibinfo{title}{{Thermal Generation of Spin Current in an
  Antiferromagnet}}.
\newblock \emph{\bibinfo{journal}{Phys. Rev. Lett.}}
  \textbf{\bibinfo{volume}{115}}, \bibinfo{pages}{266601}
  (\bibinfo{year}{2015}).

\bibitem{Wu:2016hh}
\bibinfo{author}{Wu, S.~M.} \emph{et~al.}
\newblock \bibinfo{title}{{Antiferromagnetic Spin Seebeck Effect}}.
\newblock \emph{\bibinfo{journal}{Phys. Rev. Lett.}}
  \textbf{\bibinfo{volume}{116}}, \bibinfo{pages}{097204}
  (\bibinfo{year}{2016}).

\bibitem{Wu:2015gf}
\bibinfo{author}{Wu, S.~M.}, \bibinfo{author}{Pearson, J.~E.} \&
  \bibinfo{author}{Bhattacharya, A.}
\newblock \bibinfo{title}{{Paramagnetic Spin Seebeck Effect}}.
\newblock \emph{\bibinfo{journal}{Physical Review Letters}}
  \textbf{\bibinfo{volume}{114}}, \bibinfo{pages}{186602}
  (\bibinfo{year}{2015}).

\bibitem{Hirobe:2016eg}
\bibinfo{author}{Hirobe, D.} \emph{et~al.}
\newblock \bibinfo{title}{{One-dimensional spinon spin currents}}.
\newblock \emph{\bibinfo{journal}{Nature Physics}}
  \textbf{\bibinfo{volume}{13}}, \bibinfo{pages}{30--34}
  (\bibinfo{year}{2017}).

\bibitem{Hirobe:2018ds}
\bibinfo{author}{Hirobe, D.}, \bibinfo{author}{Kawamata, T.},
  \bibinfo{author}{Oyanagi, K.}, \bibinfo{author}{Koike, Y.} \&
  \bibinfo{author}{Saitoh, E.}
\newblock \bibinfo{title}{{Generation of spin currents from one-dimensional
  quantum spin liquid}}.
\newblock \emph{\bibinfo{journal}{Journal of Applied Physics}}
  \textbf{\bibinfo{volume}{123}}, \bibinfo{pages}{123903}
  (\bibinfo{year}{2018}).

\bibitem{Pytte:1974jt}
\bibinfo{author}{Pytte, E.}
\newblock \bibinfo{title}{{Peierls instability in Heisenberg chains}}.
\newblock \emph{\bibinfo{journal}{Physical Review B}}
  \textbf{\bibinfo{volume}{10}}, \bibinfo{pages}{4637--4642}
  (\bibinfo{year}{1974}).

\bibitem{Hase:1993bo}
\bibinfo{author}{Hase, M.}, \bibinfo{author}{Terasaki, I.} \&
  \bibinfo{author}{Uchinokura, K.}
\newblock \bibinfo{title}{{Observation of the spin-Peierls transition in linear
  Cu$^{2+}$ ( Spin-$\frac{1}{2}$) chians in an inorganic compound CuGeO$_3$ }}.
\newblock \emph{\bibinfo{journal}{Physical Review Letters}}
  \textbf{\bibinfo{volume}{70}}, \bibinfo{pages}{3651--3654}
  (\bibinfo{year}{1993}).

\bibitem{2002JPCM...14R.195U}
\bibinfo{author}{Uchinokura, K.}
\newblock \bibinfo{title}{{Topical review: spin-Peierls transition in CuGeO$_3$
  and impurity-induced ordered phases in low-dimensional spin-gap systems}}.
\newblock \emph{\bibinfo{journal}{Journal of Physics: Condensed Matter}}
  \textbf{\bibinfo{volume}{14}}, \bibinfo{pages}{R195--R237}
  (\bibinfo{year}{2002}).

\bibitem{Nishi:1994da}
\bibinfo{author}{Nishi, M.}, \bibinfo{author}{Fujita, O.} \&
  \bibinfo{author}{Akimitsu, J.}
\newblock \bibinfo{title}{{Neutron-scattering study on the spin-Peierls
  transition in a quasi-one-dimensional magnet CuGeO$_3$}}.
\newblock \emph{\bibinfo{journal}{Physical Review B}}
  \textbf{\bibinfo{volume}{50}}, \bibinfo{pages}{6508--6510}
  (\bibinfo{year}{1994}).

\bibitem{Lorenzo:1994bt}
\bibinfo{author}{Lorenzo, J.~E.} \emph{et~al.}
\newblock \bibinfo{title}{{Soft longitudinal modes in spin-singlet CuGeO$_3$}}.
\newblock \emph{\bibinfo{journal}{Physical Review B}}
  \textbf{\bibinfo{volume}{50}}, \bibinfo{pages}{1278--1281}
  (\bibinfo{year}{1994}).

\bibitem{Takayoshi:2010fi}
\bibinfo{author}{Takayoshi, S.} \& \bibinfo{author}{Sato, M.}
\newblock \bibinfo{title}{{Coefficients of bosonized dimer operators in
  spin-$\frac{1}{2}$ XXZ chains and their applications}}.
\newblock \emph{\bibinfo{journal}{Physical Review B}}
  \textbf{\bibinfo{volume}{82}}, \bibinfo{pages}{214420}
  (\bibinfo{year}{2010}).

\bibitem{Brill:1994ki}
\bibinfo{author}{Brill, T.~M.} \emph{et~al.}
\newblock \bibinfo{title}{{High-Field Electron Spin Resonance and Magnetization
  in the Dimerized Phase of CuGeO$_3$}}.
\newblock \emph{\bibinfo{journal}{Physical Review Letters}}
  \textbf{\bibinfo{volume}{73}}, \bibinfo{pages}{1545} (\bibinfo{year}{1994}).

\bibitem{Regnault:1996bv}
\bibinfo{author}{Regnault, L.~P.}, \bibinfo{author}{Ain, M.},
  \bibinfo{author}{Hennion, B.}, \bibinfo{author}{Dhalenne, G.} \&
  \bibinfo{author}{Revcolevschi, A.}
\newblock \bibinfo{title}{{Inelastic-neutron-scattering investigation of the
  spin-Peierls system CuGeO$_3$}}.
\newblock \emph{\bibinfo{journal}{Physical Review B}}
  \textbf{\bibinfo{volume}{53}}, \bibinfo{pages}{5579--5597}
  (\bibinfo{year}{1996}).

\bibitem{Fujita:1995ha}
\bibinfo{author}{Fujita, O.}, \bibinfo{author}{Akimitsu, J.},
  \bibinfo{author}{Nishi, M.} \& \bibinfo{author}{Kakurai, K.}
\newblock \bibinfo{title}{{Evidence for a Singlet-Triplet Transition in
  Spin-Peierls System CuGeO$_3$}}.
\newblock \emph{\bibinfo{journal}{Physical Review Letters}}
  \textbf{\bibinfo{volume}{74}}, \bibinfo{pages}{1677} (\bibinfo{year}{1995}).

\bibitem{Bonner:1982bs}
\bibinfo{author}{Bonner, J.~C.} \& \bibinfo{author}{Bl{\"o}te, H. W.~J.}
\newblock \bibinfo{title}{{Excitation spectra of the linear alternating
  antiferromagnet}}.
\newblock \emph{\bibinfo{journal}{Physical Review B}}
  \textbf{\bibinfo{volume}{25}}, \bibinfo{pages}{6959--6980}
  (\bibinfo{year}{1982}).

\bibitem{Ronnow:2000dt}
\bibinfo{author}{R{\o}nnow, H.~M.} \emph{et~al.}
\newblock \bibinfo{title}{{Neutron Scattering Study of the Field-Induced
  Soliton Lattice in CuGeO$_3$}}.
\newblock \emph{\bibinfo{journal}{Physical Review Letters}}
  \textbf{\bibinfo{volume}{84}}, \bibinfo{pages}{4469} (\bibinfo{year}{2000}).

\bibitem{Hase:1994ez}
\bibinfo{author}{Hase, M.} \emph{et~al.}
\newblock \bibinfo{title}{{Magnetization of pure and Zn-doped spin-Peierls
  cuprate CuGeO$_3$ in high magnetic field}}.
\newblock \emph{\bibinfo{journal}{Physica B: Condensed Matter}}
  \textbf{\bibinfo{volume}{201}}, \bibinfo{pages}{167--170}
  (\bibinfo{year}{1994}).

\bibitem{Uchinokura:1995cu}
\bibinfo{author}{Uchinokura, K.}, \bibinfo{author}{Hase, M.} \&
  \bibinfo{author}{Sasago, Y.}
\newblock \bibinfo{title}{{Magnetic phase transitions in CuGeO$_3$ in high
  magnetic fields}}.
\newblock \emph{\bibinfo{journal}{Physica B: Condensed Matter}}
  \textbf{\bibinfo{volume}{211}}, \bibinfo{pages}{175--179}
  (\bibinfo{year}{1995}).

\bibitem{maekawa2012spin}
\bibinfo{author}{Maekawa, S.}, \bibinfo{author}{Valenzuela, S.},
  \bibinfo{author}{Saitoh, E.} \& \bibinfo{author}{Kimura, T.}
\newblock \emph{\bibinfo{title}{Spin Current}} (\bibinfo{publisher}{OUP Oxford,
  Oxford}, \bibinfo{year}{2012}).

\bibitem{Kikkawa:2015bn}
\bibinfo{author}{Kikkawa, T.} \emph{et~al.}
\newblock \bibinfo{title}{{Critical suppression of spin Seebeck effect by
  magnetic fields}}.
\newblock \emph{\bibinfo{journal}{Physical Review B}}
  \textbf{\bibinfo{volume}{92}}, \bibinfo{pages}{064413}
  (\bibinfo{year}{2015}).

\bibitem{2010ApPhL..97q2505U}
\bibinfo{author}{Uchida, K.-i.} \emph{et~al.}
\newblock \bibinfo{title}{{Observation of longitudinal spin-Seebeck effect in
  magnetic insulators}}.
\newblock \emph{\bibinfo{journal}{Applied Physics Letters}}
  \textbf{\bibinfo{volume}{97}}, \bibinfo{pages}{172505}
  (\bibinfo{year}{2010}).

\bibitem{Hirobe:2019cf}
\bibinfo{author}{Hirobe, D.} \emph{et~al.}
\newblock \bibinfo{title}{{Magnon Pairs and Spin-Nematic Correlation in the
  Spin Seebeck Effect}}.
\newblock \emph{\bibinfo{journal}{Physical Review Letters}}
  \textbf{\bibinfo{volume}{123}}, \bibinfo{pages}{117202}
  (\bibinfo{year}{2019}).

\bibitem{Lussier:1996gja}
\bibinfo{author}{Lussier, J.~G.}, \bibinfo{author}{Coad, S.~M.},
  \bibinfo{author}{McMorrow, D.~F.} \& \bibinfo{author}{Paul, D.~M.}
\newblock \bibinfo{title}{{The temperature dependence of the spin - Peierls
  energy gap in CuGeO$_3$}}.
\newblock \emph{\bibinfo{journal}{Journal of Physics: Condensed Matter}}
  \textbf{\bibinfo{volume}{8}}, \bibinfo{pages}{L59--L64}
  (\bibinfo{year}{1996}).

\bibitem{Sasago:1996is}
\bibinfo{author}{Sasago, Y.} \emph{et~al.}
\newblock \bibinfo{title}{{New phase diagram of Zn-doped CuGeO$_3$}}.
\newblock \emph{\bibinfo{journal}{Physical Review B}}
  \textbf{\bibinfo{volume}{54}}, \bibinfo{pages}{R6835--R6837}
  (\bibinfo{year}{1996}).

\bibitem{Fukuyama:1996hb}
\bibinfo{author}{Fukuyama, H.}, \bibinfo{author}{Tanimoto, T.} \&
  \bibinfo{author}{Saito, M.}
\newblock \bibinfo{title}{{Antiferromagnetic Long Range Order in Disordered
  Spin-Peierls Systems}}.
\newblock \emph{\bibinfo{journal}{Journal of the Physical Society of Japan}}
  \textbf{\bibinfo{volume}{65}}, \bibinfo{pages}{1182--1185}
  (\bibinfo{year}{1996}).

\bibitem{Grenier:1998ga}
\bibinfo{author}{Grenier, B.} \emph{et~al.}
\newblock \bibinfo{title}{{Magnetic susceptibility and phase diagram of
  CuGe$_{1-x}$Si$_x$O$_3$ single crystals}}.
\newblock \emph{\bibinfo{journal}{Physical Review B}}
  \textbf{\bibinfo{volume}{57}}, \bibinfo{pages}{3444--3453}
  (\bibinfo{year}{1998}).

\bibitem{abrikosov2017fundamentals}
\bibinfo{author}{Abrikosov, A.}
\newblock \emph{\bibinfo{title}{Fundamentals of the Theory of Metals}}
  (\bibinfo{publisher}{Dover Publications, Mineola, New York},
  \bibinfo{year}{2017}).

\bibitem{Adachi:2011hh}
\bibinfo{author}{Adachi, H.}, \bibinfo{author}{Ohe, J.-i.},
  \bibinfo{author}{Takahashi, S.} \& \bibinfo{author}{Maekawa, S.}
\newblock \bibinfo{title}{{Linear-response theory of spin Seebeck effect in
  ferromagnetic insulators}}.
\newblock \emph{\bibinfo{journal}{Physical Review B}}
  \textbf{\bibinfo{volume}{83}}, \bibinfo{pages}{533} (\bibinfo{year}{2011}).

\bibitem{Jauho:1994cg}
\bibinfo{author}{Jauho, A.-P.}, \bibinfo{author}{Wingreen, N.~S.} \&
  \bibinfo{author}{Meir, Y.}
\newblock \bibinfo{title}{{Time-dependent transport in interacting and
  noninteracting resonant-tunneling systems}}.
\newblock \emph{\bibinfo{journal}{Physical Review B}}
  \textbf{\bibinfo{volume}{50}}, \bibinfo{pages}{5528} (\bibinfo{year}{1994}).

\bibitem{Oyanagi:2019cv}
\bibinfo{author}{Oyanagi, K.} \emph{et~al.}
\newblock \bibinfo{title}{{Spin transport in insulators without exchange
  stiffness}}.
\newblock \emph{\bibinfo{journal}{Nature Communications}}
  \textbf{\bibinfo{volume}{10}}, \bibinfo{pages}{4740} (\bibinfo{year}{2019}).

\bibitem{Shiomi:2014gd}
\bibinfo{author}{Shiomi, Y.} \& \bibinfo{author}{Saitoh, E.}
\newblock \bibinfo{title}{{Paramagnetic Spin Pumping}}.
\newblock \emph{\bibinfo{journal}{Physical Review Letters}}
  \textbf{\bibinfo{volume}{113}}, \bibinfo{pages}{266602}
  (\bibinfo{year}{2014}).

\bibitem{Kamimura:1994fe}
\bibinfo{author}{Kamimura, O.}, \bibinfo{author}{Terauchi, M.},
  \bibinfo{author}{Tanaka, M.}, \bibinfo{author}{Fujita, O.} \&
  \bibinfo{author}{Akimitsu, J.}
\newblock \bibinfo{title}{{Electron Diffraction Study of an Inorganic
  Spin-Peierls System CuGeO$_3$}}.
\newblock \emph{\bibinfo{journal}{Journal of the Physical Society of Japan}}
  \textbf{\bibinfo{volume}{63}}, \bibinfo{pages}{2467--2471}
  (\bibinfo{year}{1994}).

\bibitem{Azuma:1994hf}
\bibinfo{author}{Azuma, M.}, \bibinfo{author}{Hiroi, Z.},
  \bibinfo{author}{Takano, M.}, \bibinfo{author}{Ishida, K.} \&
  \bibinfo{author}{Kitaoka, Y.}
\newblock \bibinfo{title}{{Observation of a Spin Gap in SrCu$_2$O$_3$
  Comprising Spin-{\textonehalf} Quasi-1D Two-Leg Ladders}}.
\newblock \emph{\bibinfo{journal}{Physical Review Letters}}
  \textbf{\bibinfo{volume}{73}}, \bibinfo{pages}{3463--3466}
  (\bibinfo{year}{1994}).

\bibitem{Kageyama:1999ke}
\bibinfo{author}{Kageyama, H.} \emph{et~al.}
\newblock \bibinfo{title}{{Exact Dimer Ground State and Quantized Magnetization
  Plateaus in the Two-Dimensional Spin System SrCu$_2$(BO$_3$)$_2$}}.
\newblock \emph{\bibinfo{journal}{Physical Review Letters}}
  \textbf{\bibinfo{volume}{82}}, \bibinfo{pages}{3168--3171}
  (\bibinfo{year}{1999}).

\end{thebibliography}
\end{document}